\documentclass[superscriptaddress,reprint,amsmath,amssymb,aps,prb,floatfix]{revtex4-2}
\usepackage{graphicx,color} 
\usepackage{dcolumn} 
\usepackage{bm,verbatim} 
\usepackage{hyperref} 
\usepackage[mathlines]{lineno} 

\newcommand{\fermiv}{v_{\text{F}}}
\newcommand{\epsp}{\varepsilon_{\bf p}}
\newcommand{\epst}{\varepsilon_{\bf p}(t)}
\newcommand{\jxpt}{\left\langle j_x\right\rangle_{p}(t)}
\newcommand{\jzpt}{\left\langle j_z\right\rangle_{p}(t)}
\newcommand{\jxt}{\left\langle j_x\right\rangle(t)}
\newcommand{\jzt}{\left\langle j_z\right\rangle(t)}
\newcommand{\npt}{n_{\bf p}(t)}
\newcommand{\eref}[1]{Eq. (\ref{#1})}
\newcommand{\figref}[1]{Fig. \ref{#1}}
\newcommand{\deff}{\Delta_{\text{eff}}}
\newcommand{\hdeff}{\hat{\Delta}_{\text{eff}}}

\begin{document}
	
	\title{Time-dependent electric transport in nodal loop semimetals}
	
	\author{Zolt\'{a}n Okv\'{a}tovity}
	\email{okvatovity.zoltan@ttk.bme.hu}
	\affiliation{Department of Theoretical Physics Budapest University of Technology and Economics, 1521 Budapest, Hungary}
	\affiliation{MTA-BME Lend\"{u}let Topology and Correlation Research Group, Budapest University of Technology and Economics, 1521 Budapest, Hungary}
	
	\author{L\'{a}szl\'{o} Oroszl\'{a}ny}
	\affiliation{Department of Physics of Complex Systems, ELTE E\"{o}tv\"{o}s Lor\'{a}nd University, 1117 Budapest, Hungary}
	\affiliation{MTA-BME Lend\"{u}let Topology and Correlation Research Group, Budapest University of Technology and Economics, 1521 Budapest, Hungary}	
	
	\author{Bal\'{a}zs D\'{o}ra}	
	\affiliation{Department of Theoretical Physics Budapest University of Technology and Economics, 1521 Budapest, Hungary}
	\affiliation{MTA-BME Lend\"{u}let Topology and Correlation Research Group, Budapest University of Technology and Economics, 1521 Budapest, Hungary}

	\date{\today}
	
	\begin{abstract}
		Close to the Fermi energy, nodal loop semimetals have a torus-shaped, strongly anisotropic Fermi surface which affects their transport properties. Here we investigate the non-equilibrium dynamics of nodal loop semimetals by going beyond linear response and determine the time evolution of the current after switching on a homogeneous electric field. The current grows monotonically with time for electric fields perpendicular to the nodal loop plane however it exhibits non-monotonical behavior for field orientations aligned within the plane. After an initial non-universal growth $\sim Et$, the current first reaches a plateau $\sim E$. Then, for perpendicular directions, it increases while for in-plane directions it decreases with time to another plateau, still $\sim E$. These features arise from interband processes.
		For long times or strong electric fields, the current grows as $\sim E^{3/2}t$ or $\sim E^3t^2$ for perpendicular or parallel electric fields, respectively.
		This non-linear response represents an intraband effect where the large number of excited quasiparticles respond to the electric field.
		Our analytical results are benchmarked by the numerical evaluation of the current from continuum and tight-binding models of nodal loop semimetals.	
	\end{abstract}
	
	\maketitle
	
	\section{\label{sec:intro}Introduction}
	
	Recently, the investigation of properties of topologically non-trivial states in solids has become one of the main focuses of condensed matter physics. The theoretical prediction\cite{Bernevig2006} and experimental realization\cite{Konig2007} of topological insulators have inspired the prediction of exotic phenomena such as the topological magnetoelectric effect\cite{Qi2009} and opened the door for exciting new applications in tools for measuring fundamental constants \cite{Maciejko_fundamental_constants_PhysRevLett.105.166803}, in thermoelectric devices\cite{xu_topological_2017}, and in architectural elements of spintronics devices\cite{brune_spin_2012}. The bulk topological insulators, just as ordinary insulators, are characterized by a gap separating the valence and conduction bands. However, in these materials, the topological properties of bulk states, characterized by the $\mathbb{Z}_2$ invariant\cite{Hasan2010}, guarantee the presence of robust, spin polarized states on the perimeter of samples.
	
	Topology can still impact the properties of systems in the absence of a band gap. The interplay of topology and symmetry can also stabilize robust features in these so-called topological semimetals \cite{Yang2018,Fang2016,Burkov2011}. In Dirac and Weyl semimetals\cite{Armitage2018,Yang2018} band degeneracies near the Fermi-level occur at a discrete set of points in the Brillouin zone. Recently these systems have been intensively studied both theoretically \cite{Vishwanath2013} and experimentally culminating in several interesting observations such as the chiral anomaly, anomalous Hall conductivity, and Fermi arc surface states\cite{Yasuoka2017,Burkov20112,Xu2015,Huang2015}.
	
	\begin{figure}[!ht]
		\includegraphics[width=\linewidth]{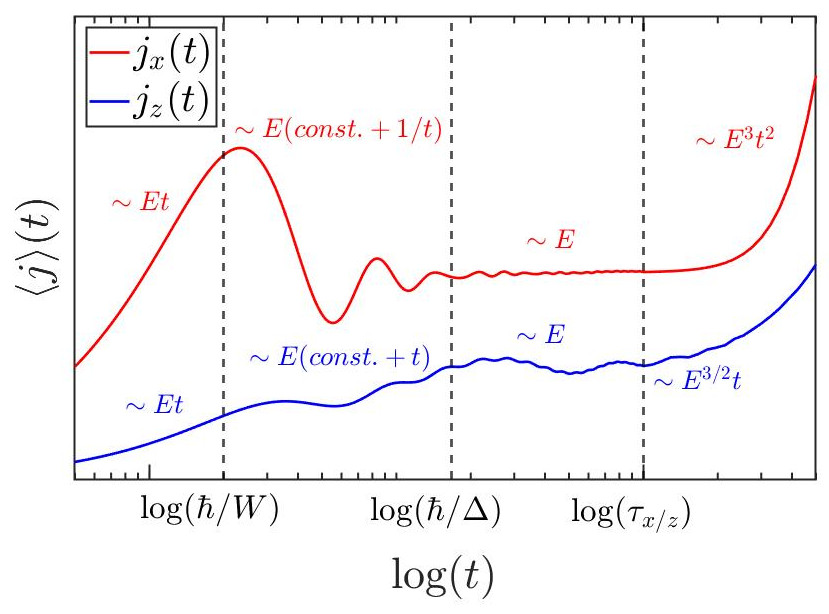}
		\caption{\label{fig:currschem} Schematic picture of the current in nodal loop semimetals when the electric field perpendicular to the nodal loop (blue line) and parallel to it (red line). The black dashed lines represents the timescales which split up the time domain into four different region.}
	\end{figure}
	
	The story of topological semimetals does not end with the Dirac and Weyl points. In certain materials, called nodal line semimetals, band crossings can appear not only in selected points but through continuous lines which may form closed loops or traverse the whole Brillouin zone\cite{Yang2018,Fang2016,Burkov2011}. Stabilizing these nodal lines against gap opening requires the presence of some additional symmetries\cite{Chiu2016,Burkov2011,Fang2015,Chiu2014}. 
	
	When the protecting symmetry is broken, either a finite band gap opens or the nodal line breaks into several nodal points in the Brillouin zone. Nodal line semimetals, in general, do not host protected edge states \cite{Fang2016}, however, localized surface states can appear between the surface projection of the nodal lines \cite{Fang2015,Chan2016} called drumhead states. These surface states owing to their dispersionless nature may provide a fertile ground for correlation-induced effects\cite{Liu2017,Le2020}.	 
	
	Model systems and material realizations of nodal line semimetals have been recently proposed in hyperhoneycomb structures\cite{Mullen2015,Ezawa2016}, superlattices made of topological insulators\cite{Burkov2011,Phillips2014}, alkaline-earth metal crystals\cite{Chan2016,Hirayama2017,Du2017} and cold atomic systems \cite{Yin2018,Zhang2016}.	
	
	There has been also intense experimental progress to investigate the surface properties using angle-resolved photoemission spectroscopy (ARPES) \cite{Kim2015,Yang2018,Bian2016} or magnetotransport experiments to reveal the bulk characteristics \cite{Hu2016,Emmanouilidou2017,Pezzini2017}. The characteristic features of nodal line semimetals have been identified in various materials e.g., PbTaSe$_2$ \cite{Bian20162,Chang2016}, ZrSiTe and ZrSiSe \cite{Hu2016,Muechler2020} or Ca$_3$P$_2$ \cite{Chan2016}. Although the lack of protected edge states makes nodal line semimetals quite challenging to identify experimentally the peculiar nature of their Fermi surface endows them with characteristic electronic and magnetic properties\cite{Tommy2020,Oroszlny2018,MartnRuiz2018}.
	
	In this paper, we investigate the non-equilibrium dynamics beyond the linear response of nodal loop semimetals after a sudden switch of a homogeneous electric field. We consider a simple continuum model with a single nodal loop located in the $(p_x,p_y)$ plane in the momentum space at the Fermi level. Following the works in Refs. \cite{Dra2010,Vajna2015}, we provide a detailed derivation of the temporal behavior of the electric current for two distinct cases: when the electric field is perpendicular to the plane of the nodal loop ($z$ direction) or parallel to it ($x$ direction). The short time/weak electric field limit of the current is obtained using first-order perturbation theory in the electric field while for the long time/strong electric field limit, we use the Dykhne--Davis--Pechukas (DDP) method. The main results are displayed in \figref{fig:currschem}. We find that in both cases, the time domain splits into four different regions. In the ultrashort regime, the current is linear in time and electric field, and the slope is determined by the high-energy cutoff. In the second region, the current in the $z$ direction reaches a plateau then starts to increase linearly while in the $x$ direction $1/t$ decay is observed. For the third region, both currents are constant yet again. This tendency of the current agrees with the suggested behavior from the frequency-dependent optical conductivity from Ref. \cite{Barati2017}. 
	
	In the last temporal regime, the electric field dependence of the current becomes non-linear. 
	For the current, we obtained $\sim E^{3/2}t$  dependence in $z$ direction and  $\sim E^3t^2$ in $x$ direction.  
	To test the applicability of our result, we calculated the current by solving the Schr\"{o}dinger equation numerically and also compared it with the current obtained from tight-binding calculations. The numerical results agree well with the time and electric field dependence of the current obtained from analytical calculations. 
	These non-linear features of the electric response are expected to be observable in transport measurements in nodal loop semimetals.	
	
	The paper is organized as follows: in Sec. \ref{sec:timedepham}, we briefly introduce the model Hamiltonian and the current operators. In Secs. \ref{sec:zdir} and \ref{sec:xdir}, the temporal and electric field dependence of the current is investigated when the electric field is perpendicular or parallel with the nodal loop, respectively. Then the results of the tight-binding calculation are detailed in Sec. \ref{sec:tb}. In Sec. \ref{sec:meas}, the experimental possibilities are briefly discussed and in Sec. \ref{sec:sum} our main results are summarized.
	
	\section{\label{sec:timedepham} Model Hamiltonian and observables}
	
	We consider the effective low-energy Hamiltonian of a nodal loop semimetal \cite{Oroszlny2018} as 
	\begin{equation}
		H=\left[\Delta -\frac{p^2_x+p^2_y}{2m}\right]\sigma_x+\fermiv p_z\sigma_z=P_x\sigma_x+P_z\sigma_z,
		\label{eq:ham0}
	\end{equation}
	where $\sigma_i$'s ($i=x,z$) are the Pauli matrices, $m\approx 0.1-1~m_e$ ($m_e$ is the mass of an electron)  is the effective mass \cite{Singha2017,Rudenko2020,Pezzini2017}, $\Delta\approx 0.1-1$ eV is the energy scale that defines the nodal loop radius  \cite{Fu2019,Hosen2017} and $\fermiv\approx 10^5-10^6$ m/s is the Fermi-velocity in the $z$ direction \cite{Singha2017,Rudenko2020}. Diagonalizing the Hamiltonian yields the energy spectrum as $E_{\pm}({\bf p})=\pm\epsp$ with $\pm$ the band index and $\epsp=\sqrt{(\fermiv p_z)^2+(\Delta-(p^2_x+p^2_y)/2m)^2}$. The homogeneous electric field switched on at $t=0$ is introduced as a time-dependent vector potential $\textbf{A}(t)$ through the Peierls substitution: $\textbf{p}\rightarrow \textbf{p}-e\textbf{A}(t)$ at $t=0$. We are interested in two different cases, when the electric field points to $x$ and $z$ direction which leads us to two different vector potentials $\textbf{A}_x(t)=[Et\Theta(t),0,0]$ and $\textbf{A}_z(t)=[0,0,Et\Theta(t)]$, respectively.
	
	For each momentum $\bf{p}$ the Hamiltonian \eref{eq:ham0} with the time-dependent vector potential	represents a two-level system. Depending on the orientation of the electric field, the instantaneous spectrum exhibits one or two (avoided) level crossings, thus a distinct temporal behavior of the electric current in the $x$ and $z$ directions is expected.
	We  investigate the current using the framework of the Landau--Zener dynamics \cite{Vitanov1996}, with the general time-dependent Schr\"{o}dinger equation given by
	\begin{eqnarray}
		H(t)&=&P_x(t)\sigma_x+P_z(t)\sigma_z,
		\label{eq:hamt}\\
		i\hbar\partial_t\Psi_{p}(t)&=&H(t)\Psi_{p}(t).
		\label{eq:sch}
	\end{eqnarray}
	It is convenient to perform a time-dependent unitary transformation first \cite{Cohen2008}, which diagonalizes $H(t)$, and brings us to the adiabatic basis \cite{Vitanov1996}. In the resulting equation, the positive and negative energy eigenstates are readily distinguished, simplifying further analytic and numerical analysis. The unitary transformation is given by
	\begin{equation}
		U=\begin{bmatrix}
			\cos\left(\frac{\theta_t}{2}\right)& \sin\left(\frac{\theta_t}{2}\right) \\
			\sin\left(\frac{\theta_t}{2}\right)& -\cos\left(\frac{\theta_t}{2}\right)
		\end{bmatrix},
		\label{eq:unitaryt}
	\end{equation}
	where $\tan(\theta_t)=P_x(t)/P_z(t)$. In the adiabatic basis, the Schr\"{o}dinger equation takes the form 
	\begin{equation}
		i\hbar\partial_t\Phi_p(t)=\left[\epst\sigma_z+F(t)\sigma_y\right]\Phi_p(t),
		\label{eq:schtrans}
	\end{equation}
	where $\Psi_{p}(t)=U\Phi_p(t)$ and $F(t)\sigma_y=-i\hbar U^{+}\partial_t U$ arise due to the explicit 
	time-dependence of the unitary transformation. $F(t)$ is referred to as diabatic coupling and is written as \cite{Lehto2012}	
	\begin{equation}
		F(t)=\hbar\frac{P_z(t)\partial_t P_x(t)-P_x(t)\partial_tP_z(t)}{2\varepsilon^2_{\bf p}(t)}.
		\label{eq:offdiagt}
	\end{equation}
	The initial condition of \eref{eq:schtrans} corresponds to half filling at zero temperature: $\Phi^T_p(t=0)=\left[0,1\right]$. 
	The current operator for a given momentum in the original basis is defined as 
	$j_p(t)=\partial H(t)/\partial {\bf A}(t)$ giving $j_x(t)=\frac{e}{m}(p_x-eEt)\sigma_x$ and $j_z(t)=-e\fermiv\sigma_z$. 
	In the adiabatic basis, the expression for the current operator is
	\begin{eqnarray}
		j_x(t)&=&\dfrac{e}{m}(p_x-eEt)\left(\sin(\theta_t)\sigma_z-\cos(\theta_t)\sigma_x\right),
		\label{eq:jxopt}\\
		j_z(t)&=&-e\fermiv\left(\cos(\theta_t)\sigma_z+\sin(\theta_t)\sigma_x\right).
		\label{eq:jzopt}
	\end{eqnarray}
	The contribution to the current from a given momentum mode is the expectation value of these operators. 
	By denoting $\Phi^T_p(t)=\left[\alpha(t),\beta(t)\right]$, we introduce the transition probability as $\npt=\left|\alpha(t)\right|^2$ which gives the number of electrons excited from the lower to the upper band (and also number of the holes remaining in the lower band). The contribution of a given momentum state to the current expressed with the transition probability reads as
	\begin{eqnarray}
		\jxpt&=&\dfrac{e}{m}\left(p_x-eEt\right)\sin(\theta_t)\left(2\npt-1\right)\nonumber\\
		&+&\dfrac{2\epst}{E}\partial_t\npt
		\label{eq:jxpt}\\
		\jzpt&=&-e\fermiv\cos(\theta_t)\left(2\npt-1\right)\nonumber\\
		&+&\dfrac{2\epst}{E}\partial_t\npt
		\label{eq:jzpt}
	\end{eqnarray}
	In both cases, the current consists of an intraband (first term) and interband (second term) part 
	which are also called conduction and polarization current in QED terminology, respectively \cite{Tanji2009,Dra2010,Vajna2015}. The total current is given by the momentum integral of the momentum resolved contributions. We note that the $\npt$ independent terms, corresponding to fully occupied or empty states, give no contribution to the current and hence can be omitted. \cite{Dra2010,Ashcroft1976}. The properties of $\npt$ and the electric current are discussed in the following sections.
	
	\section{\label{sec:zdir} Current in the z direction}
	
\begin{figure}[!ht]
		\includegraphics[width=\linewidth]{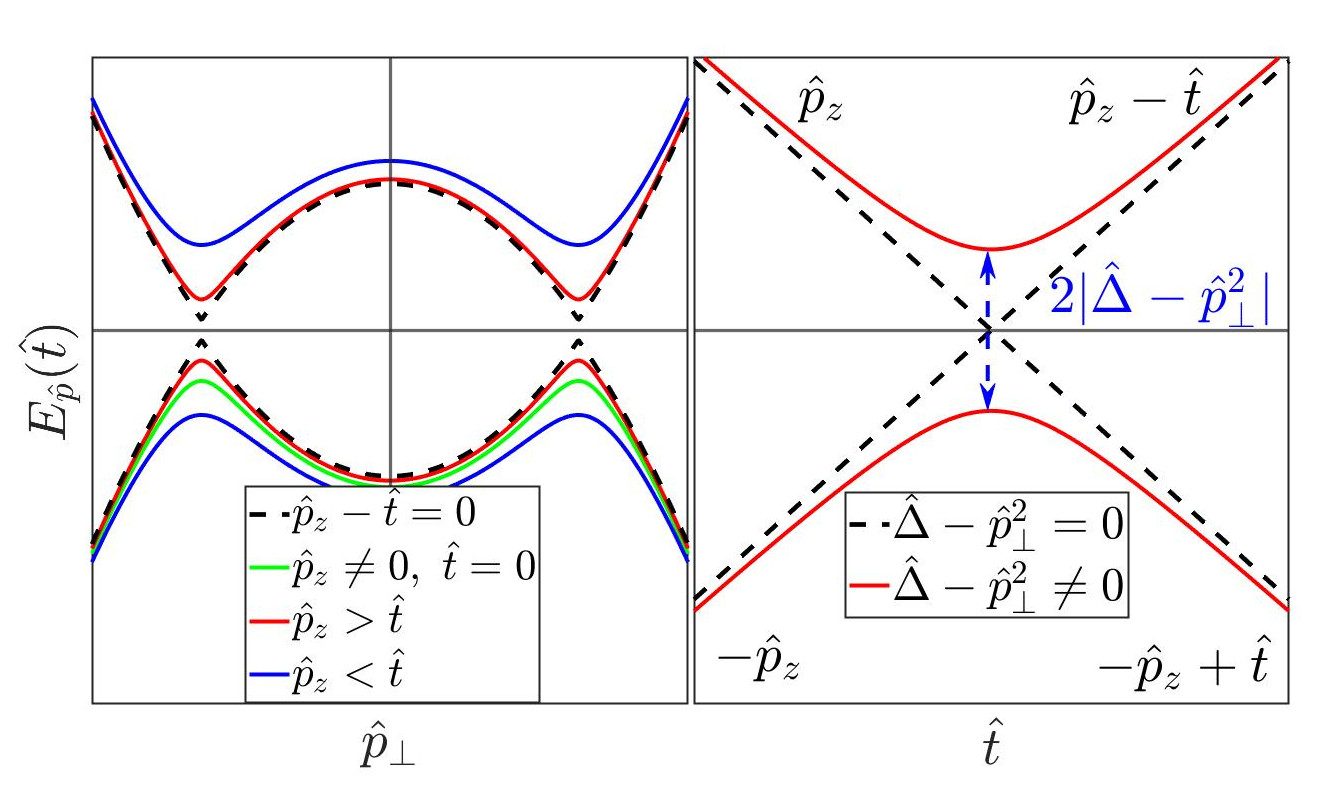}
		\caption{\label{fig:dispz} Visualization of the Landau--Zener dynamics when the $E$ points in the $z$ direction. The gap between the two bands is determined by $\hat{p}_z$ initially. During the time evolution, the gap starts to decrease and when $\hat{p}_z-\hat{t}=0$ it closes at $\hat{p}_{\perp}=\pm\sqrt{\hat{\Delta}}$, then start to increase. During the time evolution, the system is driven through a quantum critical point  \cite{Dra2010}.}
	\end{figure}

	In this section, the constant electric field is aligned to the $z$ direction i.e., it is perpendicular to the nodal loop. 
	The time-dependent vector potential is $\textbf{A}(t) = [0, 0, Et\Theta(t)]$, and the variables in \eref{eq:hamt} are $P_x=\Delta-(p^2_x+p^2_y)/2m$ which remain time independent and $P_z(t)=\fermiv(p_z-eEt)$. The evolution of the instantaneous spectrum is plotted in \figref{fig:dispz}.
	
	The time-dependent Schr\"{o}dinger equation and the current contribution in the adiabatic basis for $t>0$ read as
	\begin{eqnarray}
		i\hbar\partial_t\Phi_p(t)&=&\left[\epst\sigma_z+\dfrac{\hbar\fermiv eE}{2\varepsilon^2_{\bf p}(t)}\left(\Delta-\dfrac{p^2_{\perp}}{2m}\right)\sigma_y\right]\Phi_p(t),\nonumber \\
		\label{eq:schadz}\\
		\jzpt&=&-e\fermiv\dfrac{\fermiv(p_z-eEt)}{\epst}\left(2\npt-1\right)\nonumber\\
		&+&\dfrac{2\epst}{E}\partial_t\npt.
		\label{eq:jadz}
	\end{eqnarray}
	Here, we introduced $p_\perp=\sqrt{p^2_x+p^2_y}$. 
	The scaling properties of the Schr\"{o}dinger equation allow us to introduce dimensionless variables as  $\hat{t}=t/\tau_z$, $\hat{\Delta}=\Delta\tau_z/\hbar, \hat{p}_z=\fermiv p_z\tau_z/\hbar$, $\hat{p}_{\perp}=p_{\perp}\sqrt{\tau_z/2m\hbar}$ where the scaling factor 
	$\tau_z=\sqrt{\hbar/e\fermiv E}$ defines the natural time scale connected to the electric field. 
	The transition probability behaves differently for $\hat{t}\ll 1 $ and $\hat{t}\gg 1$, which also defines the short- and long-time limits of the total current, respectively.
	As we show below,  for short times, the dominant contribution to the current is coming from the polarization part while in the long-time limit, the current is determined by the number of excited 
	electrons in the conduction band \cite{Dra2010,Boross2011}.
	
	\subsection{\label{sec:zdirshort} Short-time evolution of the current}
\begin{figure}[!ht]
		\includegraphics[width=\linewidth]{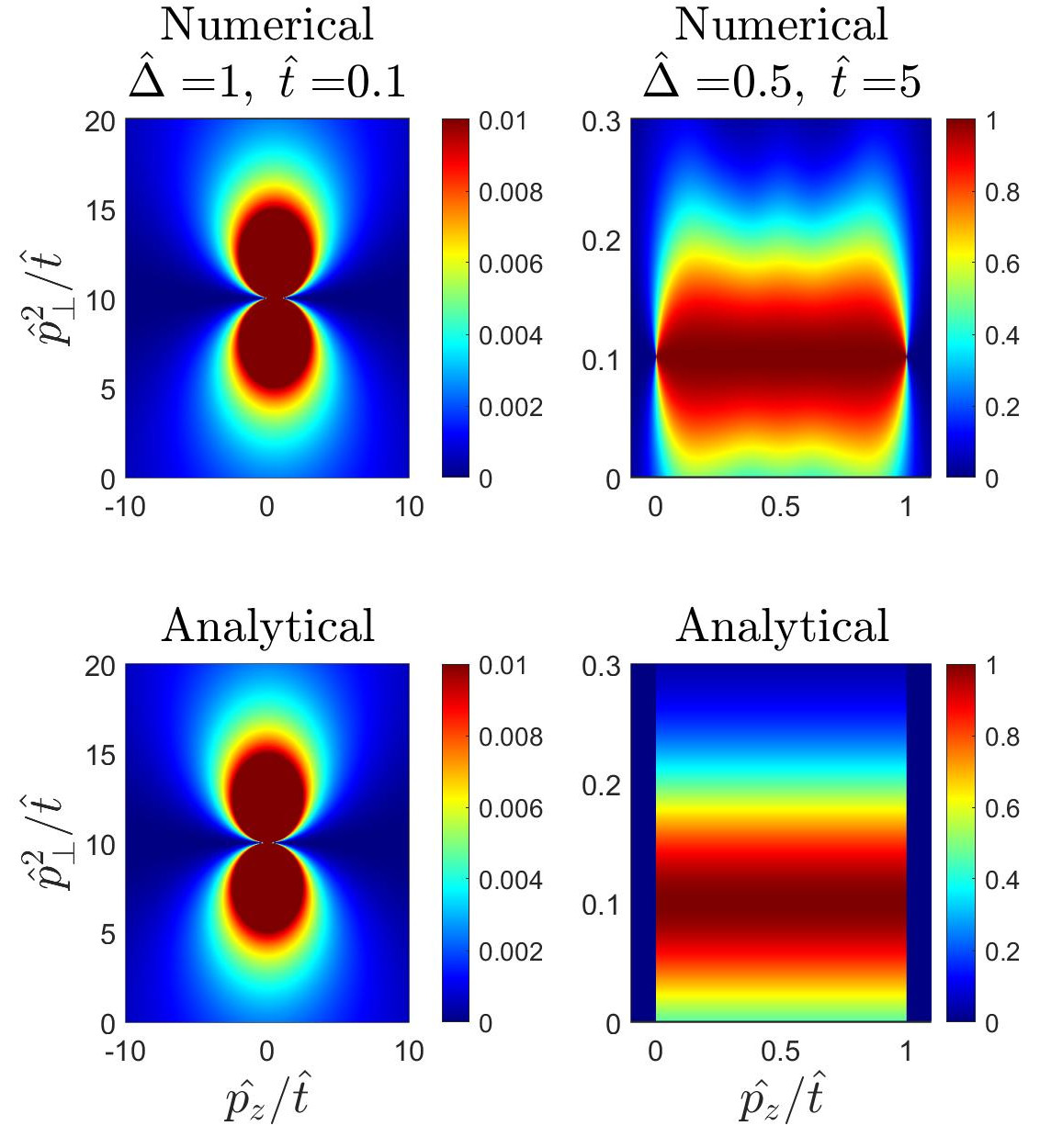}
		\caption{\label{fig:nptz} Transition probability for short (left) and long (right) times. The $\npt$ resembles closely to the transition probability obtained for graphene and Weyl semimetals \cite{Dra2010, Vajna2015} i.e., dipolar for short times and cylindrical for long times, but shifted to $\hat{p}_{\perp}=\pm\sqrt{\hat{\Delta}}$.}
	\end{figure}

	The Schr\"{o}dinger equation in \eref{eq:schadz} can be solved analytically for arbitrary times and electric fields  \cite{Tanji2009,Gavrilov1996}, but it does not give an immediately transparent solution for the transition probability. Therefore it is more practical to obtain $\npt$ from approximate solutions in different limits of $t$. 
	In the short-time limit, employing first-order perturbation theory in $E$ yields the transition probability as
	\begin{equation}
		\npt=\frac{\left(\hbar \fermiv eE\right)^2}{4\epsp^6}\left(\Delta-\frac{p^2_{\perp}}{2m}\right)^2\sin^2\left(\frac{\epsp t}{\hbar}\right)
		\label{eq:nptzs}
	\end{equation} 
	which is valid except in the close vicinity of the nodal loop i.e. $\epsp \gg \fermiv eEt$ and resembles closely to the result obtained for graphene \cite{Dra2010}. 
	To check the validity of our result, we calculated the transition probability by solving the Schr\"{o}dinger equation in \eref{eq:schadz} numerically using the explicit Runge-Kutta method. We obtained good agreement between the analytical and numerical results visualized on the left side of \figref{fig:nptz}.
	
	Using this result in \eref{eq:jadz}, the conduction part of the current contains already higher (second) order terms in electric field and gives negligible contribution to the total current. Only the polarization current contributes to the linear order in the electric field as 
	\begin{eqnarray}
		\jzt&=&\frac{\fermiv^2 e^2E}{(2\pi)^2\hbar^2}\int_{0}^{\Lambda_z}dp_z\int_{0}^{\Lambda_{\perp}}dp_{\perp}p_{\perp} \frac{(\Delta-p^2_{\perp}/2m)^2}{\epsp^4} \nonumber \\
		&\times& \sin\left(\frac{2\epsp t}{\hbar}\right),
		\label{eq:jztp}
	\end{eqnarray}
	$\Lambda_z$ and $\Lambda_{\perp}$ are the momentum cutoffs which arise from the high-energy cutoff $W$ determined by the bandwidth as $W=\fermiv \Lambda_z=\Lambda^2_{\perp}/2m$. In \eref{eq:jztp}, two different energy scales are present, $W$ and $\Delta$, which in turn determine three different temporal 
	regions: the ultrashort time transient response, when $t\ll \hbar/W$, the second when $\hbar/W\ll t\ll\hbar/\Delta$ and the third region when $\hbar/\Delta \ll t\ll \sqrt{\hbar/e\fermiv E}$. 
	
	For the ultrashort time transient response ($t\ll \hbar/W$), the current is obtained as
	\begin{equation}
		\jzpt=\frac{\fermiv^2e^2Et}{(2\pi)^2\hbar^3}\frac{(\Delta-p^2_{\perp}/2m)^2}{\sqrt{(\fermiv p_z)^2+(\Delta-p^2_{\perp}/2m)^2}^3}
		\label{eq:jzpt1}
	\end{equation} 
	The momentum integral over $p_z$ and $p_\perp$ yields
	\begin{equation}
		\jzt=\frac{me^2\fermiv E}{2\pi^2\hbar^3}Wt\ln\left[\sqrt{2}+1\right],
		\label{eq:jzt1}
	\end{equation}
	in the $\Delta\ll W$ limit. This behavior has also been observed in Dirac and Weyl fermions  \cite{Dra2010,Vajna2015} with the picture of classical particles accelerated by an external electric field. These particles obey Newton's equation with the effective mass given by $m^{-1}_{zz}=\partial^2 \epst/\partial p^2_z$.
	
\begin{figure}[!ht]
		\includegraphics[width=\linewidth]{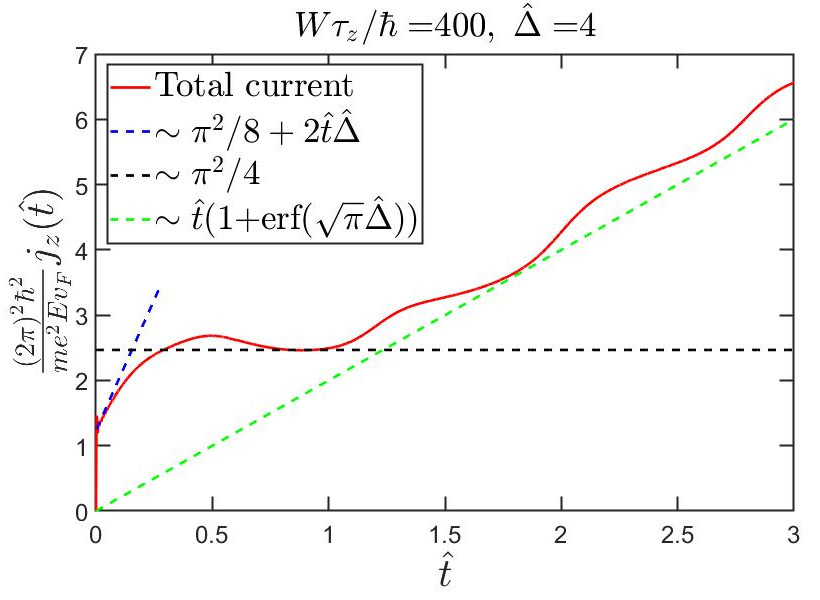}
		\caption{\label{fig:jzt} Temporal behavior of the current after switching on the electric field. 
			The blue and black dashed lines represent the polarization current in \eref{eq:jzt2} and \eref{eq:jzt3}, respectively. 
			When $\hat{t}\gg 1$ the conductive current becomes dominant, scaling as  $\sim \hat{t}$.}
	\end{figure}

	In the second region, when $\hbar/W\ll t\ll \hbar/\Delta$, the current saturates to a constant value similarly to graphene  \cite{Dra2010}. 
	This can be explained by symmetry considerations since for the electric field aligned to the $z$ direction, the cylindrical symmetry of the system remains intact. Therefore, the nodal loop can be thought of as two effective,  graphene-like systems with high-energy cutoff $W$ and $\Delta$, originating from states outside or inside the nodal loop,	respectively. Then, the current contribution coming from the first graphene-like system saturates first when $\hbar/W\ll t$ \cite{Dra2010,Vajna2015}.	With increasing time, an additional linear term in time arises from the second graphene-like system with cutoff $\Delta$. The total current can be approximated as
	\begin{equation}
		\jzt=\frac{me^2\fermiv E}{(2\pi)^2\hbar^2}\left[\frac{\pi^2}{8}+\frac{2t\Delta}{\hbar}\right].
		\label{eq:jzt2}
	\end{equation}
	In the third temporal region, the additional $\Delta$ dependent part also saturates, and the current reaches another constant value
	\begin{equation}
		\jzt=\frac{me^2\fermiv E}{16\hbar^2}.
		\label{eq:jzt3}
	\end{equation} 
	This result allows us to define a dc conductivity by taking the time-independent 
	current in \eref{eq:jzt3} and divide it with the applied electric field as $\sigma^0_z=j_z/E=me^2\fermiv/16\hbar^2$. 
	This agrees with the optical conductivity at $\omega \to 0$ in Ref. \cite{Barati2017} up to a factor of two, due to the spin degeneracy. 
	As the frequency starts to increase, the optical conductivity decreases with $1/\omega$ while for high frequencies, it tends to a constant value $\sigma^0_z/2$. All these features in the optical conductivity are in accord with our time-dependent current.
	Our analytical and numerical results agree and are illustrated in \figref{fig:jzt}.
	
	\subsection{\label{sec:zdirlong} Long-time, strong electric field limit}
	The long time, strong electric field limit, i.e., $t\gg \sqrt{\hbar/e\fermiv E}$ is out of the scope of perturbation theory. 
	Although the Schr\"{o}dinger equation can be solved exactly for arbitrary times and electric fields in terms of the parabolic cylinder functions \cite{Tanji2009}, it does not provide transparent expressions for the transition probability and the current. Instead, we rely on the so-called WKB approach to obtain  $\npt$ which is often used to determine transition probability upon tunneling through a barrier \cite{Casher1979}. In practice, we use its temporal variant, the Dykhne--Davis--Pechukas (DDP) method \cite{Suominen1992}, which is also known as Landau--Dykhne method for linear time dependence \cite{Boross2011}.
	
	For long $t$ and large $E$, the dispersion relation has linear time dependence with an (avoided) crossing visualized in \figref{fig:dispz}.
	Then using the DDP method,  we obtain for the transition probability \cite{Dra2010,Vajna2015,Boross2011}
	\begin{equation}
		\npt=\Theta(p_z)\Theta(eEt-p_z)\exp\left[-\frac{\pi(\Delta-p^2_{\perp}/2m)^2}{\hbar \fermiv eE}\right].
		\label{eq:nptzl}
	\end{equation}
	The exponential term is also called the Schwinger pair production rate \cite{Schwinger1951,Tanji2009}. The result in \eref{eq:nptzl} agrees well with transition probability obtained from numerical calculations, visualized in \figref{fig:nptz}. \eref{eq:nptzl} is applicable only when $(p_z, p_z-eEt)\gg |\Delta-p^2_{\perp}/2m|/\fermiv$ holds. 
	
	Using this, we calculate the total current, which is dominated by the conductive part  as
	\begin{equation}
		\jzt=-\frac{2 e\fermiv^2}{(2\pi)^2\hbar^3} \int_{-\infty}^{\infty}dp_z \int_{0}^{\infty}dp_\perp p_\perp\frac{p_z-eEt}{\epst}\npt.
		\label{eq:jztint}
	\end{equation}
	We can estimate the overall time and field dependence by rescaling the integral. 
	The transition between bands occurs only if the system is driven through the touching points as plotted in \figref{fig:dispz}. 
	This holds for $0\ll p_z\ll eEt$, so the number of excited electrons can be characterized by a longitudinally growing cylinder of length $\sim Et$ \cite{Vajna2015}. Using the scaling parameter $\tau_z$, we rescale $p_\perp$ as $\hat{p}_\perp=p_\perp\sqrt{\tau_z/(2m\hbar)}$ in the second integral of \eref{eq:jztint} which brings out an additional $\sim E^{1/2}$ factor. Consequently, the total current should scale with $\sim E^{3/2}t$.
	
\begin{figure}[!ht]
		\includegraphics[width=0.7\linewidth]{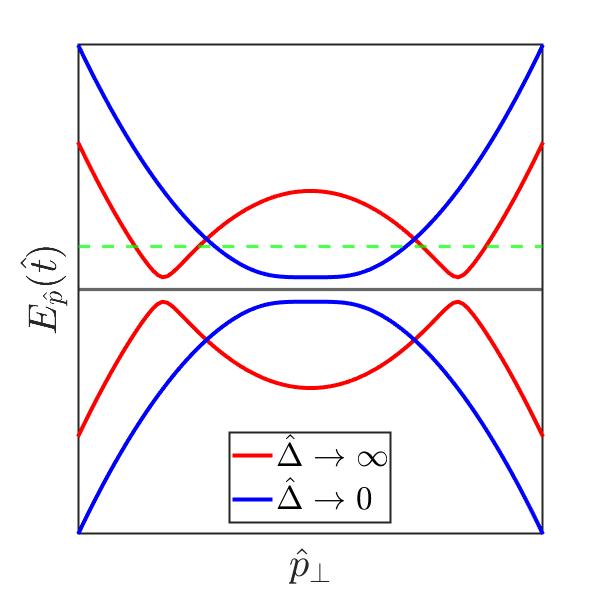}
		\caption{\label{fig:deltadep} Schematic plot of the dispersion relation in the $\hat{\Delta}\to 0$ and $\hat{\Delta}\to \infty$ limit. 
			For a given energy (green dashed line) slightly above the gap edge, there are twice as many empty states in the conduction band for $\hat{\Delta}$ large, therefore the current is $2$ times larger. }
	\end{figure}

	The integrals over $p_z$ and $p_\perp$ yield
	\begin{equation}
		\jzt = \frac{2me^2v_\text{F}E}{(2\pi)^2\hbar^2}\sqrt{\frac{v_\text{F}eE}{\hbar}}t f_z\left(\frac{\sqrt{\pi}\Delta}{\sqrt{\hbar \fermiv eE}}\right)
		\label{eq:jzt4}
	\end{equation}
	where $f_z(x)=(1+\text{erf}(x))/2$ with $\text{erf}(x)$, the error function \cite{Gradstein2014}. 
	The time and electric field dependence agree with our estimation from scaling and resembles closely to electric current in graphene \cite{Dra2010}.
	The number of excited particles is given by 
	\begin{eqnarray}
		N(t)&=&\frac{1}{(2\pi)^3\hbar^3}\int d^3p~ \npt \nonumber\\
		&=&\frac{2meE}{(2\pi)^2\hbar^2}\sqrt{\frac{v_\text{F}eE}{\hbar}}t f_z\left(\frac{\sqrt{\pi}\Delta}{\sqrt{\hbar \fermiv eE}}\right),
	\end{eqnarray}
	which leads to $\jzt=e\fermiv N(t)$. The current increases linearly with time due to the increasing number of electron-hole pairs which propagate with a constant $\fermiv$ velocity in the conduction and valence band. Due to the nodal loop, an additional $\Delta$ dependent part also arises which is responsible to a factor of $2$ enhancement in the electric current. In the $\Delta\to \infty$ or $E\to 0$ limit, due to the structure of the dispersion relation, the electrons can tunnel to twice as many states as in the $\Delta\to 0$ or $E\to \infty$ limit which explains  the factor of $2$ difference in the electric current, visualized in Fig. \ref{fig:deltadep}.
	
	\section{\label{sec:xdir} Current in the $x$ direction}
\begin{figure}[!ht]
		\includegraphics[width=\linewidth]{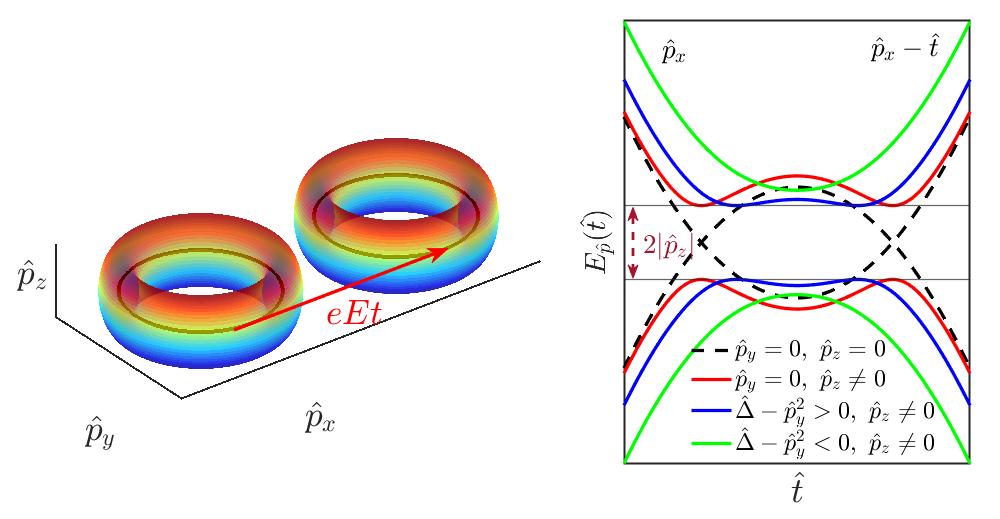}
		\caption{\label{fig:dispx} Left panel: The time evolution of the dispersion relation is illustrated. 
			The time-dependent vector potential shifts the nodal loop in the $x$ direction. Right panel: The time evolution of $\hat{p}_x$ at different values of $\hdeff$ and $\hat{p}_z$. When $\hdeff>0$, $\hat{p}_z$ determines the gap, and $\hdeff>0$ sets the location of the two minimum/maximum points. On the other hand, when $\hdeff<0$ the gap starts to increase and only one minimum/maximum remains.}
	\end{figure}

	To calculate the current in the $x$ direction, we use the vector potential $\textbf{A}(t)=[Et\Theta(t),0,0]$. 
	The variables defined in \eref{eq:hamt} are $P_z=\fermiv p_z$ and $P_x=\deff-(p_x-eEt)^2/2m$ where  $\deff=\Delta-p^2_y/2m$. 
	In contrast with the previous case, the energy-momentum dispersion relation has a $t^2$ temporal dependence for $t\to \infty$, but the nodal loop is shifted in the $x$ direction during the time evolution. The temporal behavior of the instantaneous spectrum is visualized in \figref{fig:dispx}.
	
	The time-dependent Schr\"{o}dinger equation and the electric current contribution for a given momentum mode in the adiabatic basis are given by
	\begin{eqnarray}
		i\hbar\partial_t\Phi_p(t)&=&\left[\epst\sigma_z+\frac{\hbar\fermiv eEp_z(p_x-eEt)}{2m\varepsilon^2_{\bf p}(t)}\sigma_y\right]\Phi_p(t),\nonumber\\
		\label{eq:schadx}\\
		\jxpt&=&\frac{e}{m}\frac{(p_x-eEt)}{\epst}\left(\deff-\frac{(p_x-eEt)^2}{2m}\right)\times \nonumber \\ 
		&\times&\left(2\npt-1\right)+\frac{2\epst}{E}\partial_t\npt.
		\label{eq:jadx}
	\end{eqnarray}
	We again introduce dimensionless variables as $\hat{t}=t/\tau_x$, where $\tau_x=\sqrt[3]{2m\hbar/(eE)^2}$ coming from the Schr\"{o}dinger equation, giving 
	$\hat{\Delta}_{\text{eff}}=\deff\tau_x/\hbar$, $\hat{p}_{x/y}=p_{x/y}\sqrt{\tau_x/2m\hbar}$ and $\hat{p}_z=\fermiv p_z\tau_x/\hbar$. 
	In order to analyze the Schr\"{o}dinger equation, we apply the same approximations as before, 
	namely first-order perturbation theory for $\hat{t}\ll 1$ and the DDP method for $\hat{t}\gg 1 $. 
	
	\subsection{\label{sec:xdirshort} Short-time and weak electric field limit}
	We again use lowest order perturbation theory to obtain the transition probability in the $\hat{t}\ll 1 $ limit.
	
\begin{figure}[!ht]
		\includegraphics[width=\linewidth]{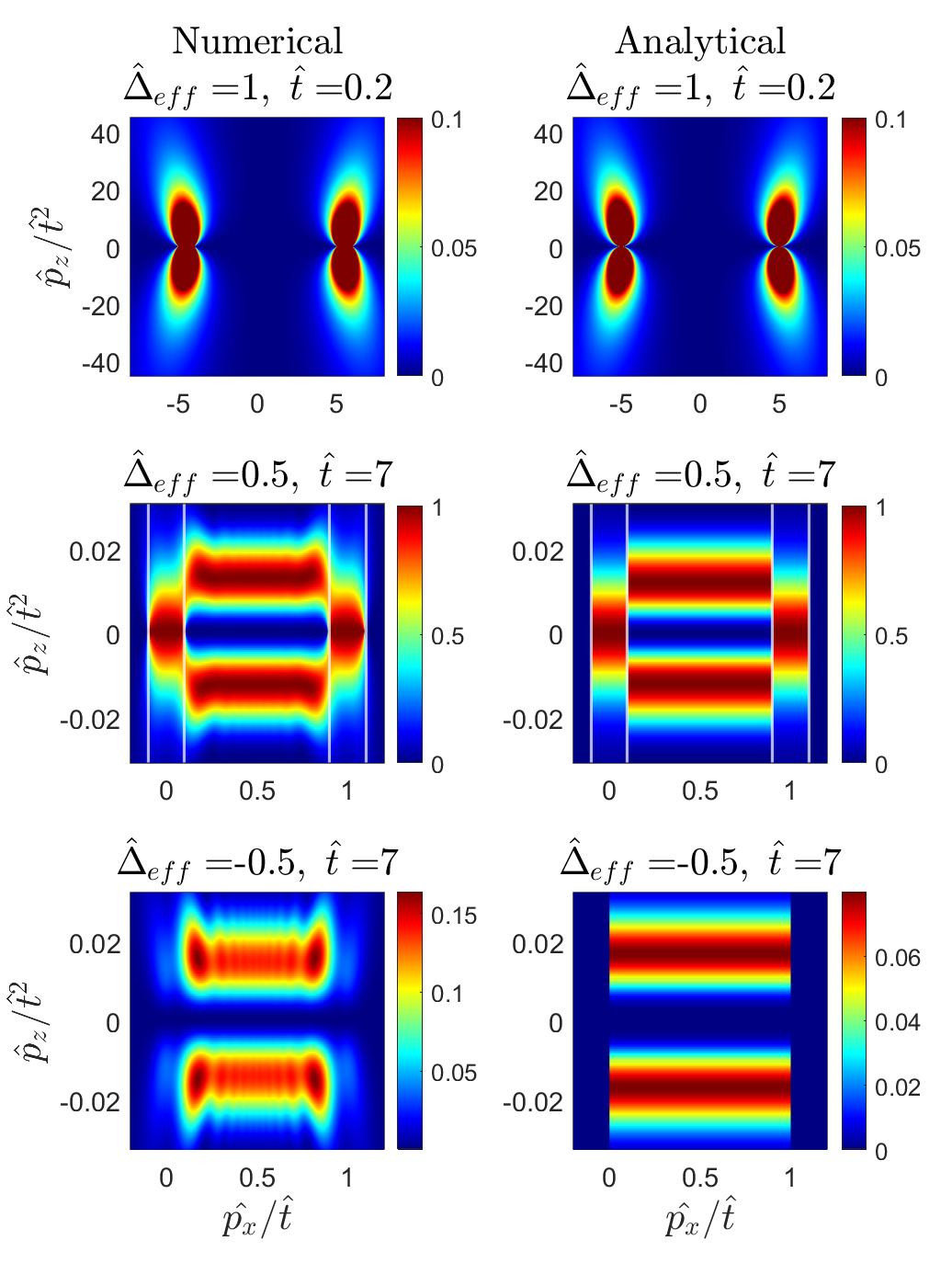}
		\caption{\label{fig:nptx} Transition probability for short (top) and long times when $\deff>0$ (middle) and $\deff<0$ (bottom). For short times, the numerical result agrees with the result from first-order perturbation theory. For long times, when $\deff>0$, we have three different regions (separated with white lines) depending on how far the nodal loop is shifted from its original position. In this case, the modified DDP formula works well for the transition probability. In the last case, when $\deff<0$ the simple DDP formula is reliable only in the adiabatic limit ($\hat{p}_z\to \infty$). 
			However, the disagreement has only a minor effect on the current since the number of excited particles decreases exponentially as $\deff\to -\infty$.}
	\end{figure}

	After some straightforward algebra, we get
	\begin{equation}
		\npt=\left(\frac{\hbar \fermiv eE}{2m}\right)^2\frac{p^2_xp^2_z}{\epsp^6}\sin^2\left(\frac{\epsp t}{\hbar}\right)
		\label{eq:nptxs}
	\end{equation}
	which is valid for $\epsp\gg (eEt)^2/2m$. The transition probability is evaluated by solving the Schr\"{o}dinger equation in \eref{eq:schadx} numerically and is plotted in \figref{fig:nptx}. For short times and  low electric fields, we obtained good agreement between the numerical and analytical result. 
	
	For short times, the dominant contribution to the current comes from the polarization term as
	\begin{eqnarray}
		\jxt&=&\frac{e^2\fermiv^2E}{2m^2(2\pi)^2\hbar^2}\int_{0}^{\Lambda_z}dp_z\int_{0}^{\Lambda_{\perp}}dp_\perp \frac{p^3_\perp p^2_z}{\epsp^4}\times\nonumber\\
		&\times&\sin\left(\frac{2\epsp t}{\hbar}\right),
		\label{eq:jxtp}
	\end{eqnarray}
	where $p_\perp=\sqrt{p^2_x+p^2_y}$ and the high-energy cutoff defined similarly to $\jzt$ as 
	$W=\fermiv \Lambda_z =\Lambda^2_{\perp}/2m$. Here again, we have two competing energy scales which separate the time domain into three distinct regions. 
	
	The ultrashort response i.e., $t\ll \hbar/W$,  is obtained by expanding the integral in time up to first order. For $W\gg\Delta$, we obtain 
	\begin{equation}
		\jxt=\frac{e^2 E t W^2}{(2\pi)^2\hbar^3\fermiv}\left[1-\sqrt{2}+\ln(1+\sqrt{2})\right] 
		\label{eq:jxt1}
	\end{equation}
	which grows linearly with time and electric field. This result is understood from fully classical consideration by applying Newton's equation with effective mass $m^{-1}_{xx}=\partial^2 \epst/\partial p^2_x$.
	For $\hbar/W\ll t\ll \hbar/\Delta$, the current does not saturate but starts to decay in time as $1/t$ and tends to a minimum value.
	In this time interval the current reads as	
	\begin{equation}
		\jxt=\frac{e^2 E}{(2\pi)^2\hbar \fermiv}\left[\frac{\Delta\pi^2}{8}+\frac{\hbar\sin^2(Wt/\hbar)}{3t}\right].
		\label{eq:jxt2}
	\end{equation}
	The oscillating part in \eref{eq:jxt2} is not universal and comes from the sharp energy cutoff. By 
	applying a smooth exponential cutoff, $\exp(-p/\Lambda)$ instead of the sharp one, oscillations are absent and the second term is modified to $2\hbar W^2t /(3(4W^2t^2+1))$. This also decays as $\sim1/t$ with increasing time, similarly to the sharp cutoff scheme which means that it is a universal characteristic feature of nodal loop semimetals. 
	
\begin{figure}[!ht]
		\includegraphics[width=\linewidth]{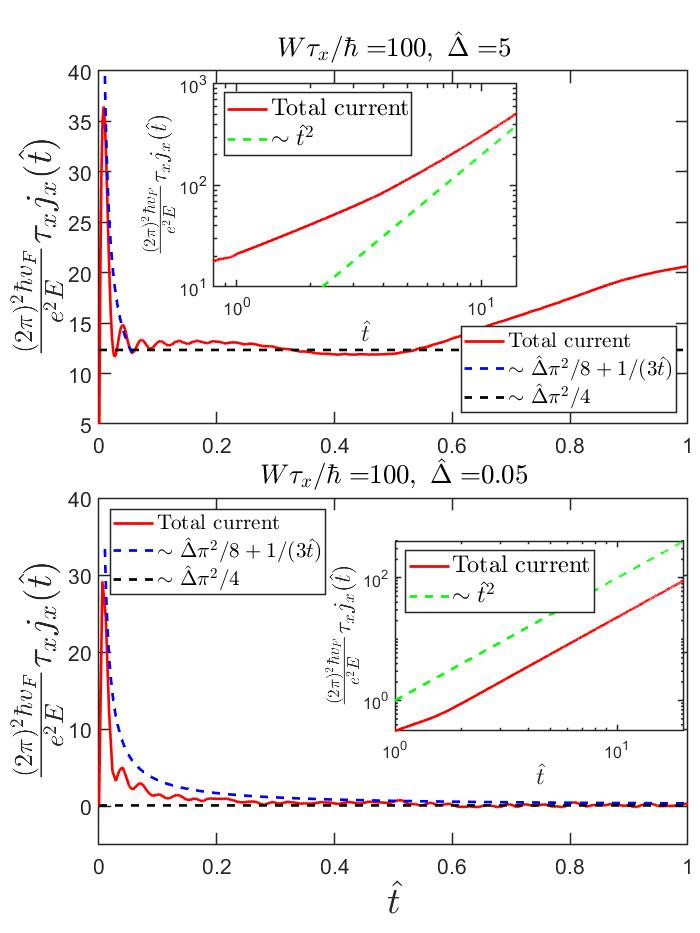}
		\caption{\label{fig:jxt} Short and long time behavior of the current after switching on the electric field in $x$ direction for large and small values of $\hat{\Delta}$. 
			The blue dashed line represents the $\sim 1/t$ decay of the polarization current while the black dashed line shows the $t \to \infty$ limit of the polarization current from \eref{eq:jxt3} which is $\sim\sigma^0_x E$. 
			The insets show the long time behavior when the conductive current is dominant. 
			For large $\hat{\Delta}$ a linear term in time can arise, due to the single channel transitions, but it is overwhelmed by the leading $\sim \hat{t}^2$  term with increasing time.}
	\end{figure}

	For $\hbar/\Delta \ll t\ll \sqrt[3]{2m\hbar/(eE)^2}$, the current tends to a time independent constant as
	\begin{equation}
		\jxt=\frac{e^2 E}{(2\pi)^2\hbar \fermiv}\left[\frac{\Delta\pi^2}{4}+\frac{\hbar(\sin^2(Wt/\hbar)-1/2)}{3t}\right].
		\label{eq:jxt3}
	\end{equation}
	Taking the $t\to \infty$ limit in \eref{eq:jxt3}, the dc response is $\sigma^0_x=j_x/E=e^2\Delta/16\hbar^2\fermiv$ which agrees with Ref. \cite{Barati2017}. Moreover, the optical conductivity grows linearly with the frequency with increasing frequency, which corresponds to the $1/t$ decay in our calculation. The current is also calculated numerically and agrees with our analytical findings in \figref{fig:jxt}.
	
	\subsection{\label{sec:xdirlong} Long-time evolution of the current}
	Once more we employ the DDP method, using the results of Ref. \cite{Suominen1992}, to elucidate the long time temporal behavior of the transition probability.
	
	For long $t$ and large $E$, the dispersion relation displays $\sim t^2$ time dependence with two (avoided) crossings, 
	visualized in \figref{fig:dispx}. The diabatic coupling reads as 
	\begin{equation}
		F({\bf p},t)=\frac{\hbar\fermiv eEp_z(p_x-eEt)}{2m\varepsilon^2_{\bf p}(t)},
		\label{eq:diab}
	\end{equation}
	which is an odd function in time for $eEt\gg p_x$ and $p_x$ is small. The wave function of \eref{eq:schadx} is rewritten
	as
	\begin{equation}
		\Phi_p(t)=
		\begin{bmatrix}
			a_1(t)\exp\left[i\int^{t}_{0}\varepsilon_{\bf p}(t') dt'\right]\\
			a_2(t)\exp\left[-i\int^{t}_{0}\varepsilon_{\bf p}(t') dt'\right]
		\end{bmatrix}
		\label{eq:phit}
	\end{equation}
	with initial conditions $a^i_1=a_1(0)=0$ and $a^i_2=a_2(0)=1$ and the transition probability is $\npt=\left|a^f_1/a^f_2\right|^2$, 
	where $a^f_{1,2}$ denotes the final states in the $t\to \infty$ limit. In the adiabatic limit, for a single crossing point the connection between the initial and final states is given by
	\begin{equation}
		\begin{bmatrix}
			a^f_1\\
			a^f_2\\
		\end{bmatrix}=
		\begin{bmatrix}
			1 & 0 \\
			e^{iD_c} & 1 
		\end{bmatrix}
		\begin{bmatrix}
			a^i_1\\
			a^i_2\\
		\end{bmatrix},
		\label{eq:s1}
	\end{equation}
	where $D_c$ is the time integral over the classically forbidden region where $\epst$ is imaginary. 
	The limits of the integral are given by the complex crossing points where $\varepsilon(t_c)=0$ as
	\begin{equation}
		t_c= \pm\frac{\sqrt{2m}}{eE}\sqrt{\deff\pm i|\fermiv p_z|},
		\label{eq:tc}
	\end{equation}
	and $D_c$ is 
	\begin{equation}
		D_c=\frac{2}{\sqrt{\beta}}\int^{\sqrt{\mu+i}}_{0}dz\sqrt{1+(\mu-z^2)^2},
		\label{eq:dcint}
	\end{equation}
	where $\beta=(\hbar eE)^2/(2m|\fermiv p_z|^3)$ and $\mu=\deff/|\fermiv p_z|$. We can identify
	$\sqrt{\beta}$ as an adiabaticity parameter since as  $\sqrt{\beta}\to 0$ the transition probability also tends to $0$  \cite{Suominen1992}. 
	The integral yields \cite{Suominen1992}
	\begin{equation}
		D_c=\frac{\pi}{2\sqrt{\beta}}\sqrt{(\mu+i)(\mu^2+1)}_2F_1\left[\frac{1}{2},-\frac{1}{2};2;\frac{\mu+i}{\mu-i}\right],
		\label{eq:dc}
	\end{equation}
	where $_2F_1(a,b;c;x)$ is the hypergeometric function\cite{Gradstein2014}. When only one channel is available to tunnel through the barrier i.e., when $\deff>0$ and $|p_x| \ll \sqrt{2m\deff}$ or  $|p_x-eEt|\ll \sqrt{2m\deff}$, the transition probability is
	\begin{equation}
		n^{1}_{\bf p}(t)=\Theta\left(\sqrt{2m\deff}-|p_x|\right)\exp\left(-2\text{Im}\left[D_c\right]\right).
		\label{eq:nptxl1}
	\end{equation}
	To take into account both tunneling channels when  $\sqrt{2m\deff}\ll p_x\ll eEt-\sqrt{2m\deff}$ for $\deff>0$, we apply the matrix from \eref{eq:s1} twice for the two crossing points, but for the second time, the off-diagonal elements pick up an extra minus sign. This extra minus sign arises due to the diabatic coupling in \eref{eq:diab}, which has a freedom in its sign \cite{Davis1976}. By fixing $F({\bf p},t)$ to be positive for the first crossing point, then we have to insert a negative sign for the off-diagonal terms since $F({\bf p},t)$ is an odd function in time for $eEt\gg p_x$ \cite{Davis1976}. The final states are given by
	\begin{equation}
		\begin{bmatrix}
			a^f_1\\
			a^f_2\\
		\end{bmatrix}=
		\begin{bmatrix}
			1 & 0 \\
			e^{iD_c}-e^{-iD^*_c} & 1 
		\end{bmatrix}
		\begin{bmatrix}
			a^i_1\\
			a^i_2\\
		\end{bmatrix},
		\label{eq:s2}
	\end{equation}
	For the two channel tunneling case, the transition probability yields
	\begin{eqnarray}
		n^{2+}_{\bf p}(t)&=&\Theta\left(p_x-\sqrt{2m\deff}\right)\Theta\left(eEt-\sqrt{2m\deff}-p_x\right) \nonumber\\
		&\times& 4\sin^2\left(\text{Re}\left[D_c\right]\right)e^{-2\text{Im}\left[D_c\right]}.
		\label{eq:nptxl2}
	\end{eqnarray} 
	This method is also applicable for the opposite, $\deff<0$ case, though with a modified time-dependent part due to the lack of the single channel transition regions. 
	In this case, the transition probability reads as
	\begin{equation}
		n^{2-}_{\bf p}(t) =\Theta\left(p_x\right)\Theta\left(eEt-p_x\right)4\sin^2\left(\text{Re}\left[D_c\right]\right)e^{-2\text{Im}\left[D_c\right]}.
		\label{eq:nptxl3}
	\end{equation} 
	These transition probabilities are expected to work in principle only in the adiabatic limit i.e., $\sqrt{\beta}\to 0$ \cite{Suominen1992,Davis1976,Dykhne1962}. 
	However, for $\deff>0$ case, we can indeed go beyond the adiabatic limit and obtain results superior to $n^{2+}_{\bf p}(t)$ in  \eref{eq:nptxl2}. For large $\deff$, we can treat our system as two, independent Landau--Zener models. Its $S$ matrix is known exactly between the initial and final states as\cite{Kazantsev1990}
	\begin{equation}
		\begin{bmatrix}
			a^f_1\\
			a^f_2\\
		\end{bmatrix}=
		\begin{bmatrix}
			\sqrt{1-|R|^2}e^{-i\chi} & -R \\
			R & \sqrt{1-|R|^2}e^{+i\chi} 
		\end{bmatrix}
		\begin{bmatrix}
			a^i_1\\
			a^i_2\\
		\end{bmatrix},
		\label{eq:s3}
	\end{equation}
	where $R=\exp[iD_c]$ with the  phase factor
	\begin{equation}
		\chi=\frac{\pi}{4}+\frac{\lambda}{2}\ln\left(\frac{\lambda}{2}\right)-\frac{\lambda}{2}+\arg\left[\Gamma\left(1-\frac{i\lambda}{2}\right)\right]
		\label{eq:chi}
	\end{equation} 
	and $\lambda=1/(2\sqrt{\beta\mu})$. Applying the $S$ matrix for both crossing points in the same way as in \eref{eq:s2}, we end up with
	\begin{eqnarray}
		n^{2\text{mod}}_{\bf p}(t)&=\nonumber&\\ &\Theta&\left(p_x-\sqrt{2m\deff}\right)\Theta\left(eEt-\sqrt{2m\deff}-p_x\right)\nonumber\\
		\times&4&\sin^2\left(\text{Re}\left[D_c\right]+\chi\right)e^{-2\text{Im}\left[D_c\right]}\left(1-e^{-2\text{Im}\left[D_c\right]}\right).\nonumber\\
		\label{eq:nptxl4}
	\end{eqnarray} 
	This expression, though similar to $n^{2+}_{\bf p}(t)$ in \eref{eq:nptxl2}, works much better and is used instead of $n^{2+}_{\bf p}(t)$ in the following.
	We compared the analytical transition probabilities from DDP and the numerical result in the second and third row of \figref{fig:nptx}. 
	For $\deff>0$, $n^{1}_{\bf p}(t)$ and $n^{2\text{mod}}_{\bf p}(t)$ agrees with the numerical calculations even beyond the adiabatic limit (i.e. $p_z\to 0$) as advertised above.
	Outside the nodal loop ($\deff <0$), the numerical and analytical results differ. However this region gives a tiny contribution to the current since as $\deff\to -\infty$ the transition probability is exponentially suppressed due to the increasing gap. Finally, the results for transition probability are summarized as 
	\begin{eqnarray}
		\npt&=&\Theta(\deff)\left(n^{1}_{\bf p}(t)+n^{1}_{eEt- p_x}(t)+n^{2\text{mod}}_{\bf p}(t)\right)\nonumber\\
		&+&\Theta(-\deff)n^{2-}_{\bf p}(t)
		\label{eq:nptxl5}
	\end{eqnarray} 	
	The temporal and electric field scaling of the current can be estimated similarly to the previous case.
	Rescaling the $p_y$ and $p_z$ variables with $\tau_x$ gives a factor of $E$. The excitation to the upper band occurs only upon complete nonadiabatic passage through the touching points which holds for $0\ll p_x\ll eEt$ so the number of excited electrons scales with $\sim Et$. 
	The velocity itself is explicitly time-dependent which also brings a factor of $\sim Et$. Combining these, we predict $E^3t^2$ scaling for the total current. 
	
	This prediction is reproduced by using \eqref{eq:nptxl4} for the conductive current contribution in \eref{eq:jadx} as
	\begin{equation}
		\jxt=-\frac{2}{(2\pi)^3\hbar^3}\int d^3p (\partial_{p_x}\epst)\npt,
		\label{eq:jxtint}
	\end{equation}
	while the polarization part gives only a subleading contribution $\sim E$.
	The one channel tunnelings contribute with $\sim t$ terms to the current for $t\to\infty$ while the other part gives the dominant, $\sim t^2$ contribution as
	\begin{equation}
		\jxt=\frac{e^4E^3}{(2\pi)^3m\hbar^2 v_{\text{F}}} t^2 f_x\left(\sqrt[3]{\frac{2m}{(\hbar eE)^2}}\Delta\right),
		\label{eq:jxt5}
	\end{equation}
	where $f_x(x)$ contains the result of $p_y$ and $p_z$ integrals and satisfies $f_x(0)=\textmd{const.}$ and $f_x(x\to\infty)\sim x^{2/3}$.
	This means that for small and large $\hat\Delta$, the current scales as $E^3t^2$ and $\Delta^{2/3}E^{23/9}t^2$, respectively. Given the large electric field exponent close to 3,
	these can be written to a good approximation as $E^3t^2$.
	
	The field and time dependence of the electric current agree with our previous estimation. The number of the excited particles are estimated as 	
	\begin{equation}
		N(t)=\int d^3p \npt\approx \frac{e^2E^2t}{(2\pi)^3\hbar v_\text{F}}f_x\left(\sqrt[3]{\frac{2m}{(\hbar eE)^2}}\Delta\right)
	\end{equation}
	which leads to $\left\langle j_x\right\rangle(t)=ev(t)N(t)$ where $v(t)=eEt/m$ is the time-dependent part of the velocity operator. Therefore, the current comes from the increasing number of particles excited to the upper band, but also these particles are accelerated by the electric field.
	
	\section{\label{sec:tb} Tight-binding model}
	
	\begin{figure}[!ht]
		\includegraphics[width=\linewidth]{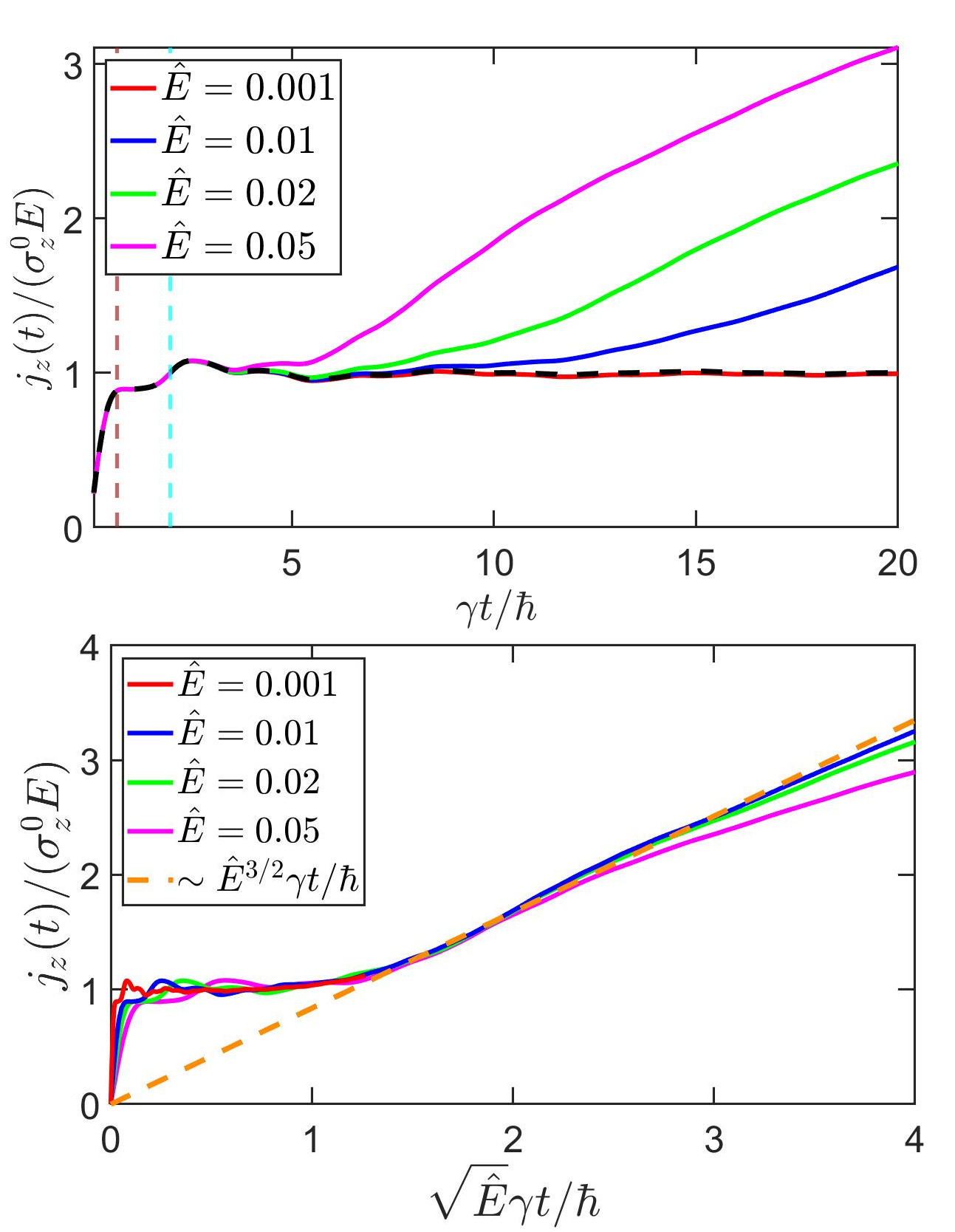}
		\caption{\label{fig:jztb1} The electric current from the tight-binding model for $\hat{E}$ in the $z$ direction. Top panel: Short-time response with $\delta/\gamma=1.5$. The brown and light blue dashed lines represent the two time scales  $1/\delta$ and $1/\Delta=1/(2\gamma-\delta)$, respectively. The polarization current dominates, represented by the black dashed line. Bottom panel: Long-time response with with $\delta/\gamma=1.5$. The orange dashed line represents the $\sim E^{3/2}t$ prediction. For large electric fields and long times, Bloch oscillations start to kick in.}
	\end{figure}
	
	\begin{figure}[!ht]
		\includegraphics[width=\linewidth]{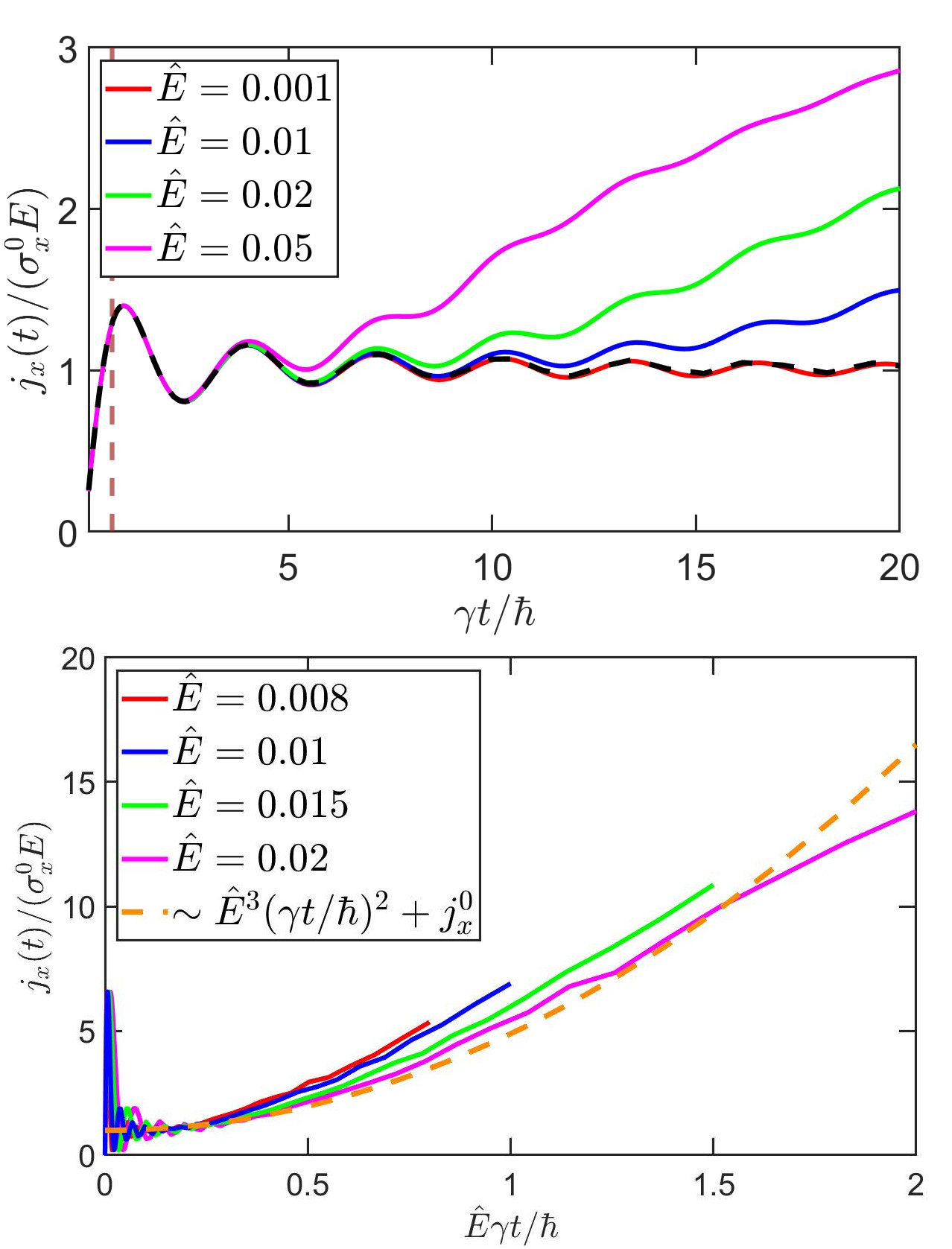}
		\caption{\label{fig:jxtb1} The electric current from the tight-binding model for $\hat{E}$  pointing to the $x$ direction. Top panel: Short-time response for $\delta/\gamma=1.5$. The brown dashed line represents the border of the ultrashort time domain at $1/\delta$. The black dashed line represents the polarization current. Bottom panel: Long-time response with $\delta/\gamma=1.97$. The orange dashed line represents our prediction $\sim E^3t^2$. The additional constant term, $j^0_x$ is the dimensionless static current coming from the polarization current. For longer times, the Bloch oscillations would kick in.}
	\end{figure}
	
	To validate the results obtained from the continuum model, we performed tight-binding calculations \cite{Zhang2016,Gu2019} based on the lattice Hamiltonian defined as
	\begin{equation}
		H_{\text{TB}}=\left[\delta -\gamma\cos(k_x a)-\gamma\cos(k_y a)\right]\sigma_x-\gamma\sin(k_z a)\sigma_z,
		\label{eq:hamtb}
	\end{equation}
	where $k_i$'s ($i=x,y,z$) are the wave numbers in different directions, $\gamma$ is the hopping integral, 
	$a$ is the lattice constant and $\delta$ determines the radius of the nodal loop. The nodal loops are located in the $k_x-k_y$ plane for $k_za=0$ and $\pi$.
	In order to avoid their overlaps and the concomitant Bloch oscillations, we use $aeEt/\hbar<\pi$ in the numerics.
	We can identify the parameters of the continuum model by expanding \eref{eq:hamtb} in $k_i$ up to second order which gives $\Delta=2\gamma-\delta$, $m=\hbar^2/a^2\gamma$ and $\fermiv=\gamma a/\hbar$. We solved the time-dependent Schr\"{o}dinger equation on a finite cubic lattice with $300$ unit cells in each direction with an adaptive grid. 
	We introduce the dimensionless electric field as $\hat{E}=eE\fermiv\hbar/\gamma^2=eE a/\gamma$. The obtained current for various electric fields are shown in Figs. \ref{fig:jztb1} and \ref{fig:jxtb1}, displaying nice agreement with our analytical findings.		
	
	In the $z$ direction, the current has two plateaus for short times, while for long times it scales with $\sim E^{3/2}t$. In the $x$ direction, the current decays for short times and grows $\sim E^{3}t^2$ for later times. Due to additional linear subleading terms scaling is adversely affected in the $x$ direction as compared to the $z$ orientation.
	
	We calculated numerically the polarization current using first-order perturbation theory in the electric field which is represented with the black dashed lines in \figref{fig:jztb1} and \figref{fig:jxtb1}. Similarly to the continuum model, the $t\to \infty$ limit defines the static current as $\sigma^0_x E$ and $\sigma^0_z E$ for $x$ and $z$ direction, respectively. Overall, the long and short-time behavior of the current from tight-binding calculations agrees well with the analytical and numerical results from the continuum model.  
	
	\section{\label{sec:meas} Experimental possibilities}
	
	So far, we discussed and evaluated the real evolution of the electric current, after switching on an electric field. Here we briefly discuss possible experiments related to measuring the current.
	
	First, by following time dependence of the current, the observation of the characteristic crossover timescales allows us to determine various parameters of nodal loop semimetals. The electric field dependent timescales $\tau_{x/z}$ separate the short- and long-time behavior. By experimentally identifying these timescales, the Fermi velocity and the effective mass can be obtained as $\fermiv=\hbar/(eE\tau^2_z)$ and $m=\tau^3_x(eE)^2/(2\hbar)$. The short-time evolution of the current also contains useful information about the nodal loop parameter $\Delta$ and the bandwidth. When the electric field aligned to the $z$ direction, two plateaus are observable when $t>\hbar/W$ and $t>\hbar/\Delta$. Identifying the temporal crossing points when the current reaches these plateaus allows us to estimate the bandwidth and the nodal loop parameter. Similarly, the peak in $j_x$ current before the $1/t$ decay also determines the bandwidth.
	
	One way to detect the nonlinear electric response is to experimentally realize these nodal loop semimetals in ultracold atomic systems \cite{Tarruell2012,Song2019}. Applying a weak magnetic-field gradient \cite{Tarruell2012} or tuning the frequency difference of laser waves responsible for the optical lattice \cite{BenDahan1996} create a constant force, which is equivalent to switching on an electric field in solid-state systems. The main advantages of these measurements are the absence of scattering and dissipation and strong electric fields are not needed to obtain the nonlinear response. According to our tight-binding calculations, the current from the electron-hole pair creation is observable before the Bloch oscillations kick in.
	
	Another way to obtain the short and late time electric response of nodal loop semimetals is to measure the current in a solid-state realization. However, in these systems scattering processes appear due to the phonons, impurities, etc. The Drude picture provides a simple way to interpret our results in the presence of impurities: The charge carriers move ballistically until a momentum transfer happens due to scattering processes. The average lifetime can be characterized by the relaxation time $\tau_{sc}$, which introduces a restricting time scale to our system. For $t>\tau_{sc}$, the current become stationary, which is estimated by substituting $t\to \tau_{sc}$ in the corresponding expressions for the current. Consequently, long-time features of the electric current are only observable if $\tau_{x/z}<\tau_{sc}$, which is equivalent to $E>E_{x/z,c}$, where $E_{x/z,c}$ is the critical electric field which separates the linear from the non-linear regions. It is defined as $E_{z,c}=\hbar/(e\fermiv \tau^2_{sc})$ for $z$ direction and $E_{x,c}=\sqrt{2m\hbar/(e^2\tau^3_{sc})}$ for $x$ direction. The scattering time is estimated as  $\tau_{sc}\sim 10^{-2}-10^{-1}$ ps \cite{Novak2019,Rudenko2020}, which implies that the minimal electric field required to observe nonlinear transport is $E_{x/z,c}\sim10^5-10^7$ V/m. For $E_E>E_{x/z,c}$, the current changes its slope as a function of the electric field, but an even larger electric field window may be required to obtain the corresponding exponents.
	
	\section{\label{sec:sum} Conclusion}
	
	\begin{table*}[t!]
		\caption{\label{tab:results} Time evolution and electric field dependence of the current in nodal loop semimetals.}
		\begin{tabular}{|l|c|c|c|c|}\hline
			& Ultrashort response & Kubo I & Kubo II & Long time response \\
			& $t\ll \hbar/W$ & $\hbar/W \ll t \ll \hbar/\Delta$ & $\hbar/\Delta \ll t \ll \tau_{x/z}$ & $\tau_{x/z} \ll t$ \\ \hline
			$x$ direction & $j_x\sim Et$ & $j_x\sim E(const. + 1/t)$ & $j_x\sim E$ & $j_x\sim E^3t^2$ \\ \hline
			$z$ direction & $j_z\sim Et$ & $j_z\sim E(const. + t)$   & $j_z\sim E$ & $j_z\sim E^{3/2}t$ \\ \hline
		\end{tabular}
	\end{table*}
	
	In this paper, we have investigated the time evolution of the non-equilibrium electric current of nodal loop semimetals after switching on a homogeneous electric field. 
	We considered the two characteristic cases, namely when the electric field is within the plane of the loop or perpendicular to it.
	To calculate the current, we determined the transition probabilities by using a variety of techniques, including first-order perturbation theory for short times and weak electric fields and the DDP method for long times and strong electric fields. Based on this, the intra- and interband contributions to the electric current are identified.
	
	For short times and weak electric fields, the interband processes dominate the current for both electric field orientations, and the ensuing time dependence can also be formally understood from a Kubo formula calculation of the optical conductivity. 
	For long times and strong fields, on the other hand, the current originates from intraband processes, namely by the increasing number of excited quasiparticles in the initially empty upper band. 
	In addition, we benchmarked our analytical results by the numerical solution of the Schr\"{o}dinger equation
	both in the continuum limit and for tight-binding models. Our results are summarized in Table \ref{tab:results}.
	
	\begin{acknowledgments}
		This research is supported by the National Research, Development and Innovation Office - NKFIH within the Quantum Information National Laboratory of Hungary and Quantum Technology National Excellence Program (Project No.	2017-1.2.1-NKP-2017-00001), K119442, K134437, K131938, FK124723, and by the Romanian National Authority for Scientific Research and Innovation, UEFISCDI, under Project No. PN-III-P4-ID-PCE-2020-0277.
		
		L. O. also acknowledges support from of the NRDI Office of Hungary and the Hungarian Academy of Sciences through the Bolyai and Bolyai+ scholarships.
	\end{acknowledgments}
	
	\bibliographystyle{apsrev}
	\bibliography{nodal_loop_current}

\begin{thebibliography}{66}
\expandafter\ifx\csname natexlab\endcsname\relax\def\natexlab#1{#1}\fi
\expandafter\ifx\csname bibnamefont\endcsname\relax
  \def\bibnamefont#1{#1}\fi
\expandafter\ifx\csname bibfnamefont\endcsname\relax
  \def\bibfnamefont#1{#1}\fi
\expandafter\ifx\csname citenamefont\endcsname\relax
  \def\citenamefont#1{#1}\fi
\expandafter\ifx\csname url\endcsname\relax
  \def\url#1{\texttt{#1}}\fi
\expandafter\ifx\csname urlprefix\endcsname\relax\def\urlprefix{URL }\fi
\providecommand{\bibinfo}[2]{#2}
\providecommand{\eprint}[2][]{\url{#2}}

\bibitem[{\citenamefont{Bernevig et~al.}(2006)\citenamefont{Bernevig, Hughes,
  and Zhang}}]{Bernevig2006}
\bibinfo{author}{\bibfnamefont{B.~A.} \bibnamefont{Bernevig}},
  \bibinfo{author}{\bibfnamefont{T.~L.} \bibnamefont{Hughes}},
  \bibnamefont{and} \bibinfo{author}{\bibfnamefont{S.-C.} \bibnamefont{Zhang}},
  \bibinfo{journal}{Science} \textbf{\bibinfo{volume}{314}},
  \bibinfo{pages}{1757} (\bibinfo{year}{2006}),
  \urlprefix\url{https://doi.org/10.1126/science.1133734}.

\bibitem[{\citenamefont{Konig et~al.}(2007)\citenamefont{Konig, Wiedmann,
  Brune, Roth, Buhmann, Molenkamp, Qi, and Zhang}}]{Konig2007}
\bibinfo{author}{\bibfnamefont{M.}~\bibnamefont{Konig}},
  \bibinfo{author}{\bibfnamefont{S.}~\bibnamefont{Wiedmann}},
  \bibinfo{author}{\bibfnamefont{C.}~\bibnamefont{Brune}},
  \bibinfo{author}{\bibfnamefont{A.}~\bibnamefont{Roth}},
  \bibinfo{author}{\bibfnamefont{H.}~\bibnamefont{Buhmann}},
  \bibinfo{author}{\bibfnamefont{L.~W.} \bibnamefont{Molenkamp}},
  \bibinfo{author}{\bibfnamefont{X.-L.} \bibnamefont{Qi}}, \bibnamefont{and}
  \bibinfo{author}{\bibfnamefont{S.-C.} \bibnamefont{Zhang}},
  \bibinfo{journal}{Science} \textbf{\bibinfo{volume}{318}},
  \bibinfo{pages}{766} (\bibinfo{year}{2007}),
  \urlprefix\url{https://doi.org/10.1126/science.1148047}.

\bibitem[{\citenamefont{Qi et~al.}(2009)\citenamefont{Qi, Li, Zang, and
  Zhang}}]{Qi2009}
\bibinfo{author}{\bibfnamefont{X.-L.} \bibnamefont{Qi}},
  \bibinfo{author}{\bibfnamefont{R.}~\bibnamefont{Li}},
  \bibinfo{author}{\bibfnamefont{J.}~\bibnamefont{Zang}}, \bibnamefont{and}
  \bibinfo{author}{\bibfnamefont{S.-C.} \bibnamefont{Zhang}},
  \bibinfo{journal}{Science} \textbf{\bibinfo{volume}{323}},
  \bibinfo{pages}{1184} (\bibinfo{year}{2009}),
  \urlprefix\url{https://doi.org/10.1126/science.1167747}.

\bibitem[{\citenamefont{Maciejko et~al.}(2010)\citenamefont{Maciejko, Qi, Drew,
  and Zhang}}]{Maciejko_fundamental_constants_PhysRevLett.105.166803}
\bibinfo{author}{\bibfnamefont{J.}~\bibnamefont{Maciejko}},
  \bibinfo{author}{\bibfnamefont{X.-L.} \bibnamefont{Qi}},
  \bibinfo{author}{\bibfnamefont{H.~D.} \bibnamefont{Drew}}, \bibnamefont{and}
  \bibinfo{author}{\bibfnamefont{S.-C.} \bibnamefont{Zhang}},
  \bibinfo{journal}{Physical Review Letters} \textbf{\bibinfo{volume}{105}},
  \bibinfo{pages}{166803} (\bibinfo{year}{2010}).

\bibitem[{\citenamefont{Xu et~al.}(2017)\citenamefont{Xu, Xu, and
  Zhu}}]{xu_topological_2017}
\bibinfo{author}{\bibfnamefont{N.}~\bibnamefont{Xu}},
  \bibinfo{author}{\bibfnamefont{Y.}~\bibnamefont{Xu}}, \bibnamefont{and}
  \bibinfo{author}{\bibfnamefont{J.}~\bibnamefont{Zhu}}, \bibinfo{journal}{npj
  Quantum Materials} \textbf{\bibinfo{volume}{2}}, \bibinfo{pages}{51}
  (\bibinfo{year}{2017}).

\bibitem[{\citenamefont{Brüne et~al.}(2012)\citenamefont{Brüne, Roth,
  Buhmann, Hankiewicz, Molenkamp, Maciejko, Qi, and Zhang}}]{brune_spin_2012}
\bibinfo{author}{\bibfnamefont{C.}~\bibnamefont{Brüne}},
  \bibinfo{author}{\bibfnamefont{A.}~\bibnamefont{Roth}},
  \bibinfo{author}{\bibfnamefont{H.}~\bibnamefont{Buhmann}},
  \bibinfo{author}{\bibfnamefont{E.~M.} \bibnamefont{Hankiewicz}},
  \bibinfo{author}{\bibfnamefont{L.~W.} \bibnamefont{Molenkamp}},
  \bibinfo{author}{\bibfnamefont{J.}~\bibnamefont{Maciejko}},
  \bibinfo{author}{\bibfnamefont{X.-L.} \bibnamefont{Qi}}, \bibnamefont{and}
  \bibinfo{author}{\bibfnamefont{S.-C.} \bibnamefont{Zhang}},
  \bibinfo{journal}{Nature Physics} \textbf{\bibinfo{volume}{8}},
  \bibinfo{pages}{485} (\bibinfo{year}{2012}), ISSN \bibinfo{issn}{1745-2481},
  \urlprefix\url{https://www.nature.com/articles/nphys2322}.

\bibitem[{\citenamefont{Hasan and Kane}(2010)}]{Hasan2010}
\bibinfo{author}{\bibfnamefont{M.~Z.} \bibnamefont{Hasan}} \bibnamefont{and}
  \bibinfo{author}{\bibfnamefont{C.~L.} \bibnamefont{Kane}},
  \bibinfo{journal}{Reviews of Modern Physics} \textbf{\bibinfo{volume}{82}},
  \bibinfo{pages}{3045} (\bibinfo{year}{2010}),
  \urlprefix\url{https://doi.org/10.1103/revmodphys.82.3045}.

\bibitem[{\citenamefont{Yang et~al.}(2018)\citenamefont{Yang, Yang, Derunova,
  Parkin, Yan, and Ali}}]{Yang2018}
\bibinfo{author}{\bibfnamefont{S.-Y.} \bibnamefont{Yang}},
  \bibinfo{author}{\bibfnamefont{H.}~\bibnamefont{Yang}},
  \bibinfo{author}{\bibfnamefont{E.}~\bibnamefont{Derunova}},
  \bibinfo{author}{\bibfnamefont{S.~S.~P.} \bibnamefont{Parkin}},
  \bibinfo{author}{\bibfnamefont{B.}~\bibnamefont{Yan}}, \bibnamefont{and}
  \bibinfo{author}{\bibfnamefont{M.~N.} \bibnamefont{Ali}},
  \bibinfo{journal}{Advances in Physics: X} \textbf{\bibinfo{volume}{3}},
  \bibinfo{pages}{1414631} (\bibinfo{year}{2018}),
  \urlprefix\url{https://doi.org/10.1080/23746149.2017.1414631}.

\bibitem[{\citenamefont{Fang et~al.}(2016)\citenamefont{Fang, Weng, Dai, and
  Fang}}]{Fang2016}
\bibinfo{author}{\bibfnamefont{C.}~\bibnamefont{Fang}},
  \bibinfo{author}{\bibfnamefont{H.}~\bibnamefont{Weng}},
  \bibinfo{author}{\bibfnamefont{X.}~\bibnamefont{Dai}}, \bibnamefont{and}
  \bibinfo{author}{\bibfnamefont{Z.}~\bibnamefont{Fang}},
  \bibinfo{journal}{Chinese Physics B} \textbf{\bibinfo{volume}{25}},
  \bibinfo{pages}{117106} (\bibinfo{year}{2016}),
  \urlprefix\url{https://doi.org/10.1088/1674-1056/25/11/117106}.

\bibitem[{\citenamefont{Burkov et~al.}(2011)\citenamefont{Burkov, Hook, and
  Balents}}]{Burkov2011}
\bibinfo{author}{\bibfnamefont{A.~A.} \bibnamefont{Burkov}},
  \bibinfo{author}{\bibfnamefont{M.~D.} \bibnamefont{Hook}}, \bibnamefont{and}
  \bibinfo{author}{\bibfnamefont{L.}~\bibnamefont{Balents}},
  \bibinfo{journal}{Physical Review B} \textbf{\bibinfo{volume}{84}},
  \bibinfo{pages}{235126} (\bibinfo{year}{2011}),
  \urlprefix\url{https://doi.org/10.1103/physrevb.84.235126}.

\bibitem[{\citenamefont{Armitage et~al.}(2018)\citenamefont{Armitage, Mele, and
  Vishwanath}}]{Armitage2018}
\bibinfo{author}{\bibfnamefont{N.~P.} \bibnamefont{Armitage}},
  \bibinfo{author}{\bibfnamefont{E.~J.} \bibnamefont{Mele}}, \bibnamefont{and}
  \bibinfo{author}{\bibfnamefont{A.}~\bibnamefont{Vishwanath}},
  \bibinfo{journal}{Reviews of Modern Physics} \textbf{\bibinfo{volume}{90}},
  \bibinfo{pages}{015001} (\bibinfo{year}{2018}),
  \urlprefix\url{https://doi.org/10.1103/revmodphys.90.015001}.

\bibitem[{\citenamefont{Turner and Vishwanath}(2013)}]{Vishwanath2013}
\bibinfo{author}{\bibfnamefont{A.}~\bibnamefont{Turner}} \bibnamefont{and}
  \bibinfo{author}{\bibfnamefont{A.}~\bibnamefont{Vishwanath}},
  \emph{\bibinfo{title}{Topological Insulators: Chapter 11. Beyond Band
  Insulators: Topology of Semimetals and Interacting Phases}}, Contemporary
  Concepts of Condensed Matter Science (\bibinfo{publisher}{Elsevier Science},
  \bibinfo{year}{2013}), ISBN \bibinfo{isbn}{9780128086926}.

\bibitem[{\citenamefont{Yasuoka et~al.}(2017)\citenamefont{Yasuoka, Kubo,
  Kishimoto, Kasinathan, Schmidt, Yan, Zhang, Tou, Felser, Mackenzie
  et~al.}}]{Yasuoka2017}
\bibinfo{author}{\bibfnamefont{H.}~\bibnamefont{Yasuoka}},
  \bibinfo{author}{\bibfnamefont{T.}~\bibnamefont{Kubo}},
  \bibinfo{author}{\bibfnamefont{Y.}~\bibnamefont{Kishimoto}},
  \bibinfo{author}{\bibfnamefont{D.}~\bibnamefont{Kasinathan}},
  \bibinfo{author}{\bibfnamefont{M.}~\bibnamefont{Schmidt}},
  \bibinfo{author}{\bibfnamefont{B.}~\bibnamefont{Yan}},
  \bibinfo{author}{\bibfnamefont{Y.}~\bibnamefont{Zhang}},
  \bibinfo{author}{\bibfnamefont{H.}~\bibnamefont{Tou}},
  \bibinfo{author}{\bibfnamefont{C.}~\bibnamefont{Felser}},
  \bibinfo{author}{\bibfnamefont{A.~P.} \bibnamefont{Mackenzie}},
  \bibnamefont{et~al.}, \bibinfo{journal}{Physical Review Letters}
  \textbf{\bibinfo{volume}{118}}, \bibinfo{pages}{236403}
  (\bibinfo{year}{2017}),
  \urlprefix\url{https://doi.org/10.1103/physrevlett.118.236403}.

\bibitem[{\citenamefont{Burkov and Balents}(2011)}]{Burkov20112}
\bibinfo{author}{\bibfnamefont{A.~A.} \bibnamefont{Burkov}} \bibnamefont{and}
  \bibinfo{author}{\bibfnamefont{L.}~\bibnamefont{Balents}},
  \bibinfo{journal}{Physical Review Letters} \textbf{\bibinfo{volume}{107}},
  \bibinfo{pages}{127205} (\bibinfo{year}{2011}),
  \urlprefix\url{https://doi.org/10.1103/physrevlett.107.127205}.

\bibitem[{\citenamefont{Xu et~al.}(2015)\citenamefont{Xu, Belopolski, Alidoust,
  Neupane, Bian, Zhang, Sankar, Chang, Yuan, Lee et~al.}}]{Xu2015}
\bibinfo{author}{\bibfnamefont{S.-Y.} \bibnamefont{Xu}},
  \bibinfo{author}{\bibfnamefont{I.}~\bibnamefont{Belopolski}},
  \bibinfo{author}{\bibfnamefont{N.}~\bibnamefont{Alidoust}},
  \bibinfo{author}{\bibfnamefont{M.}~\bibnamefont{Neupane}},
  \bibinfo{author}{\bibfnamefont{G.}~\bibnamefont{Bian}},
  \bibinfo{author}{\bibfnamefont{C.}~\bibnamefont{Zhang}},
  \bibinfo{author}{\bibfnamefont{R.}~\bibnamefont{Sankar}},
  \bibinfo{author}{\bibfnamefont{G.}~\bibnamefont{Chang}},
  \bibinfo{author}{\bibfnamefont{Z.}~\bibnamefont{Yuan}},
  \bibinfo{author}{\bibfnamefont{C.-C.} \bibnamefont{Lee}},
  \bibnamefont{et~al.}, \bibinfo{journal}{Science}
  \textbf{\bibinfo{volume}{349}}, \bibinfo{pages}{613} (\bibinfo{year}{2015}),
  \urlprefix\url{https://doi.org/10.1126/science.aaa9297}.

\bibitem[{\citenamefont{Huang et~al.}(2015)\citenamefont{Huang, Xu, Belopolski,
  Lee, Chang, Wang, Alidoust, Bian, Neupane, Zhang et~al.}}]{Huang2015}
\bibinfo{author}{\bibfnamefont{S.-M.} \bibnamefont{Huang}},
  \bibinfo{author}{\bibfnamefont{S.-Y.} \bibnamefont{Xu}},
  \bibinfo{author}{\bibfnamefont{I.}~\bibnamefont{Belopolski}},
  \bibinfo{author}{\bibfnamefont{C.-C.} \bibnamefont{Lee}},
  \bibinfo{author}{\bibfnamefont{G.}~\bibnamefont{Chang}},
  \bibinfo{author}{\bibfnamefont{B.}~\bibnamefont{Wang}},
  \bibinfo{author}{\bibfnamefont{N.}~\bibnamefont{Alidoust}},
  \bibinfo{author}{\bibfnamefont{G.}~\bibnamefont{Bian}},
  \bibinfo{author}{\bibfnamefont{M.}~\bibnamefont{Neupane}},
  \bibinfo{author}{\bibfnamefont{C.}~\bibnamefont{Zhang}},
  \bibnamefont{et~al.}, \bibinfo{journal}{Nature Communications}
  \textbf{\bibinfo{volume}{6}}, \bibinfo{pages}{8373} (\bibinfo{year}{2015}),
  \urlprefix\url{https://doi.org/10.1038/ncomms8373}.

\bibitem[{\citenamefont{Chiu et~al.}(2016)\citenamefont{Chiu, Teo, Schnyder,
  and Ryu}}]{Chiu2016}
\bibinfo{author}{\bibfnamefont{C.-K.} \bibnamefont{Chiu}},
  \bibinfo{author}{\bibfnamefont{J.~C.~Y.} \bibnamefont{Teo}},
  \bibinfo{author}{\bibfnamefont{A.~P.} \bibnamefont{Schnyder}},
  \bibnamefont{and} \bibinfo{author}{\bibfnamefont{S.}~\bibnamefont{Ryu}},
  \bibinfo{journal}{Reviews of Modern Physics} \textbf{\bibinfo{volume}{88}},
  \bibinfo{pages}{035005} (\bibinfo{year}{2016}),
  \urlprefix\url{https://doi.org/10.1103/revmodphys.88.035005}.

\bibitem[{\citenamefont{Fang et~al.}(2015)\citenamefont{Fang, Chen, Kee, and
  Fu}}]{Fang2015}
\bibinfo{author}{\bibfnamefont{C.}~\bibnamefont{Fang}},
  \bibinfo{author}{\bibfnamefont{Y.}~\bibnamefont{Chen}},
  \bibinfo{author}{\bibfnamefont{H.-Y.} \bibnamefont{Kee}}, \bibnamefont{and}
  \bibinfo{author}{\bibfnamefont{L.}~\bibnamefont{Fu}},
  \bibinfo{journal}{Physical Review B} \textbf{\bibinfo{volume}{92}},
  \bibinfo{pages}{081201(R)} (\bibinfo{year}{2015}),
  \urlprefix\url{https://doi.org/10.1103/physrevb.92.081201}.

\bibitem[{\citenamefont{Chiu and Schnyder}(2014)}]{Chiu2014}
\bibinfo{author}{\bibfnamefont{C.-K.} \bibnamefont{Chiu}} \bibnamefont{and}
  \bibinfo{author}{\bibfnamefont{A.~P.} \bibnamefont{Schnyder}},
  \bibinfo{journal}{Physical Review B} \textbf{\bibinfo{volume}{90}},
  \bibinfo{pages}{205136} (\bibinfo{year}{2014}),
  \urlprefix\url{https://doi.org/10.1103/physrevb.90.205136}.

\bibitem[{\citenamefont{Chan et~al.}(2016)\citenamefont{Chan, Chiu, Chou, and
  Schnyder}}]{Chan2016}
\bibinfo{author}{\bibfnamefont{Y.-H.} \bibnamefont{Chan}},
  \bibinfo{author}{\bibfnamefont{C.-K.} \bibnamefont{Chiu}},
  \bibinfo{author}{\bibfnamefont{M.~Y.} \bibnamefont{Chou}}, \bibnamefont{and}
  \bibinfo{author}{\bibfnamefont{A.~P.} \bibnamefont{Schnyder}},
  \bibinfo{journal}{Physical Review B} \textbf{\bibinfo{volume}{93}},
  \bibinfo{pages}{205132} (\bibinfo{year}{2016}),
  \urlprefix\url{https://doi.org/10.1103/physrevb.93.205132}.

\bibitem[{\citenamefont{Liu and Balents}(2017)}]{Liu2017}
\bibinfo{author}{\bibfnamefont{J.}~\bibnamefont{Liu}} \bibnamefont{and}
  \bibinfo{author}{\bibfnamefont{L.}~\bibnamefont{Balents}},
  \bibinfo{journal}{Physical Review B} \textbf{\bibinfo{volume}{95}},
  \bibinfo{pages}{075426} (\bibinfo{year}{2017}),
  \urlprefix\url{https://doi.org/10.1103/physrevb.95.075426}.

\bibitem[{\citenamefont{Le et~al.}(2020)\citenamefont{Le, Sun, Jin, Che, Yin,
  Li, Pang, Xu, Zhao, Kittaka et~al.}}]{Le2020}
\bibinfo{author}{\bibfnamefont{T.}~\bibnamefont{Le}},
  \bibinfo{author}{\bibfnamefont{Y.}~\bibnamefont{Sun}},
  \bibinfo{author}{\bibfnamefont{H.-K.} \bibnamefont{Jin}},
  \bibinfo{author}{\bibfnamefont{L.}~\bibnamefont{Che}},
  \bibinfo{author}{\bibfnamefont{L.}~\bibnamefont{Yin}},
  \bibinfo{author}{\bibfnamefont{J.}~\bibnamefont{Li}},
  \bibinfo{author}{\bibfnamefont{G.}~\bibnamefont{Pang}},
  \bibinfo{author}{\bibfnamefont{C.}~\bibnamefont{Xu}},
  \bibinfo{author}{\bibfnamefont{L.}~\bibnamefont{Zhao}},
  \bibinfo{author}{\bibfnamefont{S.}~\bibnamefont{Kittaka}},
  \bibnamefont{et~al.}, \bibinfo{journal}{Science Bulletin}
  \textbf{\bibinfo{volume}{65}}, \bibinfo{pages}{1349} (\bibinfo{year}{2020}),
  \urlprefix\url{https://doi.org/10.1016/j.scib.2020.04.039}.

\bibitem[{\citenamefont{Mullen et~al.}(2015)\citenamefont{Mullen, Uchoa, and
  Glatzhofer}}]{Mullen2015}
\bibinfo{author}{\bibfnamefont{K.}~\bibnamefont{Mullen}},
  \bibinfo{author}{\bibfnamefont{B.}~\bibnamefont{Uchoa}}, \bibnamefont{and}
  \bibinfo{author}{\bibfnamefont{D.~T.} \bibnamefont{Glatzhofer}},
  \bibinfo{journal}{Physical Review Letters} \textbf{\bibinfo{volume}{115}},
  \bibinfo{pages}{026403} (\bibinfo{year}{2015}),
  \urlprefix\url{https://doi.org/10.1103/physrevlett.115.026403}.

\bibitem[{\citenamefont{Ezawa}(2016)}]{Ezawa2016}
\bibinfo{author}{\bibfnamefont{M.}~\bibnamefont{Ezawa}},
  \bibinfo{journal}{Physical Review Letters} \textbf{\bibinfo{volume}{116}},
  \bibinfo{pages}{127202} (\bibinfo{year}{2016}),
  \urlprefix\url{https://doi.org/10.1103/physrevlett.116.127202}.

\bibitem[{\citenamefont{Phillips and Aji}(2014)}]{Phillips2014}
\bibinfo{author}{\bibfnamefont{M.}~\bibnamefont{Phillips}} \bibnamefont{and}
  \bibinfo{author}{\bibfnamefont{V.}~\bibnamefont{Aji}},
  \bibinfo{journal}{Physical Review B} \textbf{\bibinfo{volume}{90}},
  \bibinfo{pages}{115111} (\bibinfo{year}{2014}),
  \urlprefix\url{https://doi.org/10.1103/physrevb.90.115111}.

\bibitem[{\citenamefont{Hirayama et~al.}(2017)\citenamefont{Hirayama, Okugawa,
  Miyake, and Murakami}}]{Hirayama2017}
\bibinfo{author}{\bibfnamefont{M.}~\bibnamefont{Hirayama}},
  \bibinfo{author}{\bibfnamefont{R.}~\bibnamefont{Okugawa}},
  \bibinfo{author}{\bibfnamefont{T.}~\bibnamefont{Miyake}}, \bibnamefont{and}
  \bibinfo{author}{\bibfnamefont{S.}~\bibnamefont{Murakami}},
  \bibinfo{journal}{Nature Communications} \textbf{\bibinfo{volume}{8}},
  \bibinfo{pages}{14022} (\bibinfo{year}{2017}),
  \urlprefix\url{https://doi.org/10.1038/ncomms14022}.

\bibitem[{\citenamefont{Du et~al.}(2017)\citenamefont{Du, Tang, Wang, Sheng,
  jun Kan, Duan, Savrasov, and Wan}}]{Du2017}
\bibinfo{author}{\bibfnamefont{Y.}~\bibnamefont{Du}},
  \bibinfo{author}{\bibfnamefont{F.}~\bibnamefont{Tang}},
  \bibinfo{author}{\bibfnamefont{D.}~\bibnamefont{Wang}},
  \bibinfo{author}{\bibfnamefont{L.}~\bibnamefont{Sheng}},
  \bibinfo{author}{\bibfnamefont{E.}~\bibnamefont{jun Kan}},
  \bibinfo{author}{\bibfnamefont{C.-G.} \bibnamefont{Duan}},
  \bibinfo{author}{\bibfnamefont{S.~Y.} \bibnamefont{Savrasov}},
  \bibnamefont{and} \bibinfo{author}{\bibfnamefont{X.}~\bibnamefont{Wan}},
  \bibinfo{journal}{npj Quantum Materials} \textbf{\bibinfo{volume}{2}},
  \bibinfo{pages}{41535} (\bibinfo{year}{2017}),
  \urlprefix\url{https://doi.org/10.1038/s41535-016-0005-4}.

\bibitem[{\citenamefont{Yin et~al.}(2018)\citenamefont{Yin, Li, and
  Chen}}]{Yin2018}
\bibinfo{author}{\bibfnamefont{C.}~\bibnamefont{Yin}},
  \bibinfo{author}{\bibfnamefont{L.}~\bibnamefont{Li}}, \bibnamefont{and}
  \bibinfo{author}{\bibfnamefont{S.}~\bibnamefont{Chen}},
  \bibinfo{journal}{Physical Review A} \textbf{\bibinfo{volume}{97}},
  \bibinfo{pages}{013604} (\bibinfo{year}{2018}),
  \urlprefix\url{https://doi.org/10.1103/physreva.97.013604}.

\bibitem[{\citenamefont{Zhang et~al.}(2016)\citenamefont{Zhang, Zhao, Liu, Xue,
  Zhu, and Wang}}]{Zhang2016}
\bibinfo{author}{\bibfnamefont{D.-W.} \bibnamefont{Zhang}},
  \bibinfo{author}{\bibfnamefont{Y.~X.} \bibnamefont{Zhao}},
  \bibinfo{author}{\bibfnamefont{R.-B.} \bibnamefont{Liu}},
  \bibinfo{author}{\bibfnamefont{Z.-Y.} \bibnamefont{Xue}},
  \bibinfo{author}{\bibfnamefont{S.-L.} \bibnamefont{Zhu}}, \bibnamefont{and}
  \bibinfo{author}{\bibfnamefont{Z.~D.} \bibnamefont{Wang}},
  \bibinfo{journal}{Physical Review A} \textbf{\bibinfo{volume}{93}},
  \bibinfo{pages}{043617} (\bibinfo{year}{2016}),
  \urlprefix\url{https://doi.org/10.1103/physreva.93.043617}.

\bibitem[{\citenamefont{Kim et~al.}(2015)\citenamefont{Kim, Wieder, Kane, and
  Rappe}}]{Kim2015}
\bibinfo{author}{\bibfnamefont{Y.}~\bibnamefont{Kim}},
  \bibinfo{author}{\bibfnamefont{B.~J.} \bibnamefont{Wieder}},
  \bibinfo{author}{\bibfnamefont{C.~L.} \bibnamefont{Kane}}, \bibnamefont{and}
  \bibinfo{author}{\bibfnamefont{A.~M.} \bibnamefont{Rappe}},
  \bibinfo{journal}{Physical Review Letters} \textbf{\bibinfo{volume}{115}},
  \bibinfo{pages}{036806} (\bibinfo{year}{2015}),
  \urlprefix\url{https://doi.org/10.1103/physrevlett.115.036806}.

\bibitem[{\citenamefont{Bian et~al.}(2016{\natexlab{a}})\citenamefont{Bian,
  Chang, Zheng, Velury, Xu, Neupert, Chiu, Huang, Sanchez, Belopolski
  et~al.}}]{Bian2016}
\bibinfo{author}{\bibfnamefont{G.}~\bibnamefont{Bian}},
  \bibinfo{author}{\bibfnamefont{T.-R.} \bibnamefont{Chang}},
  \bibinfo{author}{\bibfnamefont{H.}~\bibnamefont{Zheng}},
  \bibinfo{author}{\bibfnamefont{S.}~\bibnamefont{Velury}},
  \bibinfo{author}{\bibfnamefont{S.-Y.} \bibnamefont{Xu}},
  \bibinfo{author}{\bibfnamefont{T.}~\bibnamefont{Neupert}},
  \bibinfo{author}{\bibfnamefont{C.-K.} \bibnamefont{Chiu}},
  \bibinfo{author}{\bibfnamefont{S.-M.} \bibnamefont{Huang}},
  \bibinfo{author}{\bibfnamefont{D.~S.} \bibnamefont{Sanchez}},
  \bibinfo{author}{\bibfnamefont{I.}~\bibnamefont{Belopolski}},
  \bibnamefont{et~al.}, \bibinfo{journal}{Physical Review B}
  \textbf{\bibinfo{volume}{93}}, \bibinfo{pages}{121113(R)}
  (\bibinfo{year}{2016}{\natexlab{a}}),
  \urlprefix\url{https://doi.org/10.1103/physrevb.93.121113}.

\bibitem[{\citenamefont{Hu et~al.}(2016)\citenamefont{Hu, Tang, Liu, Liu, Zhu,
  Graf, Myhro, Tran, Lau, Wei et~al.}}]{Hu2016}
\bibinfo{author}{\bibfnamefont{J.}~\bibnamefont{Hu}},
  \bibinfo{author}{\bibfnamefont{Z.}~\bibnamefont{Tang}},
  \bibinfo{author}{\bibfnamefont{J.}~\bibnamefont{Liu}},
  \bibinfo{author}{\bibfnamefont{X.}~\bibnamefont{Liu}},
  \bibinfo{author}{\bibfnamefont{Y.}~\bibnamefont{Zhu}},
  \bibinfo{author}{\bibfnamefont{D.}~\bibnamefont{Graf}},
  \bibinfo{author}{\bibfnamefont{K.}~\bibnamefont{Myhro}},
  \bibinfo{author}{\bibfnamefont{S.}~\bibnamefont{Tran}},
  \bibinfo{author}{\bibfnamefont{C.~N.} \bibnamefont{Lau}},
  \bibinfo{author}{\bibfnamefont{J.}~\bibnamefont{Wei}}, \bibnamefont{et~al.},
  \bibinfo{journal}{Physical Review Letters} \textbf{\bibinfo{volume}{117}},
  \bibinfo{pages}{016602} (\bibinfo{year}{2016}),
  \urlprefix\url{https://doi.org/10.1103/physrevlett.117.016602}.

\bibitem[{\citenamefont{Emmanouilidou et~al.}(2017)\citenamefont{Emmanouilidou,
  Shen, Deng, Chang, Shi, Kotliar, Xu, and Ni}}]{Emmanouilidou2017}
\bibinfo{author}{\bibfnamefont{E.}~\bibnamefont{Emmanouilidou}},
  \bibinfo{author}{\bibfnamefont{B.}~\bibnamefont{Shen}},
  \bibinfo{author}{\bibfnamefont{X.}~\bibnamefont{Deng}},
  \bibinfo{author}{\bibfnamefont{T.-R.} \bibnamefont{Chang}},
  \bibinfo{author}{\bibfnamefont{A.}~\bibnamefont{Shi}},
  \bibinfo{author}{\bibfnamefont{G.}~\bibnamefont{Kotliar}},
  \bibinfo{author}{\bibfnamefont{S.-Y.} \bibnamefont{Xu}}, \bibnamefont{and}
  \bibinfo{author}{\bibfnamefont{N.}~\bibnamefont{Ni}},
  \bibinfo{journal}{Physical Review B} \textbf{\bibinfo{volume}{95}},
  \bibinfo{pages}{245113} (\bibinfo{year}{2017}),
  \urlprefix\url{https://doi.org/10.1103/physrevb.95.245113}.

\bibitem[{\citenamefont{Pezzini et~al.}(2017)\citenamefont{Pezzini, van Delft,
  Schoop, Lotsch, Carrington, Katsnelson, Hussey, and Wiedmann}}]{Pezzini2017}
\bibinfo{author}{\bibfnamefont{S.}~\bibnamefont{Pezzini}},
  \bibinfo{author}{\bibfnamefont{M.~R.} \bibnamefont{van Delft}},
  \bibinfo{author}{\bibfnamefont{L.~M.} \bibnamefont{Schoop}},
  \bibinfo{author}{\bibfnamefont{B.~V.} \bibnamefont{Lotsch}},
  \bibinfo{author}{\bibfnamefont{A.}~\bibnamefont{Carrington}},
  \bibinfo{author}{\bibfnamefont{M.~I.} \bibnamefont{Katsnelson}},
  \bibinfo{author}{\bibfnamefont{N.~E.} \bibnamefont{Hussey}},
  \bibnamefont{and} \bibinfo{author}{\bibfnamefont{S.}~\bibnamefont{Wiedmann}},
  \bibinfo{journal}{Nature Physics} \textbf{\bibinfo{volume}{14}},
  \bibinfo{pages}{178} (\bibinfo{year}{2017}),
  \urlprefix\url{https://doi.org/10.1038/nphys4306}.

\bibitem[{\citenamefont{Bian et~al.}(2016{\natexlab{b}})\citenamefont{Bian,
  Chang, Sankar, Xu, Zheng, Neupert, Chiu, Huang, Chang, Belopolski
  et~al.}}]{Bian20162}
\bibinfo{author}{\bibfnamefont{G.}~\bibnamefont{Bian}},
  \bibinfo{author}{\bibfnamefont{T.-R.} \bibnamefont{Chang}},
  \bibinfo{author}{\bibfnamefont{R.}~\bibnamefont{Sankar}},
  \bibinfo{author}{\bibfnamefont{S.-Y.} \bibnamefont{Xu}},
  \bibinfo{author}{\bibfnamefont{H.}~\bibnamefont{Zheng}},
  \bibinfo{author}{\bibfnamefont{T.}~\bibnamefont{Neupert}},
  \bibinfo{author}{\bibfnamefont{C.-K.} \bibnamefont{Chiu}},
  \bibinfo{author}{\bibfnamefont{S.-M.} \bibnamefont{Huang}},
  \bibinfo{author}{\bibfnamefont{G.}~\bibnamefont{Chang}},
  \bibinfo{author}{\bibfnamefont{I.}~\bibnamefont{Belopolski}},
  \bibnamefont{et~al.}, \bibinfo{journal}{Nature Communications}
  \textbf{\bibinfo{volume}{7}}, \bibinfo{pages}{10556}
  (\bibinfo{year}{2016}{\natexlab{b}}),
  \urlprefix\url{https://doi.org/10.1038/ncomms10556}.

\bibitem[{\citenamefont{Chang et~al.}(2016)\citenamefont{Chang, Chen, Bian,
  Huang, Zheng, Neupert, Sankar, Xu, Belopolski, Chang et~al.}}]{Chang2016}
\bibinfo{author}{\bibfnamefont{T.-R.} \bibnamefont{Chang}},
  \bibinfo{author}{\bibfnamefont{P.-J.} \bibnamefont{Chen}},
  \bibinfo{author}{\bibfnamefont{G.}~\bibnamefont{Bian}},
  \bibinfo{author}{\bibfnamefont{S.-M.} \bibnamefont{Huang}},
  \bibinfo{author}{\bibfnamefont{H.}~\bibnamefont{Zheng}},
  \bibinfo{author}{\bibfnamefont{T.}~\bibnamefont{Neupert}},
  \bibinfo{author}{\bibfnamefont{R.}~\bibnamefont{Sankar}},
  \bibinfo{author}{\bibfnamefont{S.-Y.} \bibnamefont{Xu}},
  \bibinfo{author}{\bibfnamefont{I.}~\bibnamefont{Belopolski}},
  \bibinfo{author}{\bibfnamefont{G.}~\bibnamefont{Chang}},
  \bibnamefont{et~al.}, \bibinfo{journal}{Physical Review B}
  \textbf{\bibinfo{volume}{93}}, \bibinfo{pages}{245130}
  (\bibinfo{year}{2016}),
  \urlprefix\url{https://doi.org/10.1103/physrevb.93.245130}.

\bibitem[{\citenamefont{Muechler et~al.}(2020)\citenamefont{Muechler, Topp,
  Queiroz, Krivenkov, Varykhalov, Cano, Ast, and Schoop}}]{Muechler2020}
\bibinfo{author}{\bibfnamefont{L.}~\bibnamefont{Muechler}},
  \bibinfo{author}{\bibfnamefont{A.}~\bibnamefont{Topp}},
  \bibinfo{author}{\bibfnamefont{R.}~\bibnamefont{Queiroz}},
  \bibinfo{author}{\bibfnamefont{M.}~\bibnamefont{Krivenkov}},
  \bibinfo{author}{\bibfnamefont{A.}~\bibnamefont{Varykhalov}},
  \bibinfo{author}{\bibfnamefont{J.}~\bibnamefont{Cano}},
  \bibinfo{author}{\bibfnamefont{C.~R.} \bibnamefont{Ast}}, \bibnamefont{and}
  \bibinfo{author}{\bibfnamefont{L.~M.} \bibnamefont{Schoop}},
  \bibinfo{journal}{Physical Review X} \textbf{\bibinfo{volume}{10}},
  \bibinfo{pages}{011026} (\bibinfo{year}{2020}),
  \urlprefix\url{https://doi.org/10.1103/physrevx.10.011026}.

\bibitem[{\citenamefont{Tai and Lee}(2021)}]{Tommy2020}
\bibinfo{author}{\bibfnamefont{T.}~\bibnamefont{Tai}} \bibnamefont{and}
  \bibinfo{author}{\bibfnamefont{C.~H.} \bibnamefont{Lee}},
  \bibinfo{journal}{Physical Review B} \textbf{\bibinfo{volume}{103}},
  \bibinfo{pages}{195125} (\bibinfo{year}{2021}),
  \urlprefix\url{https://doi.org/10.1103/physrevb.103.195125}.

\bibitem[{\citenamefont{Oroszl{\'{a}}ny
  et~al.}(2018)\citenamefont{Oroszl{\'{a}}ny, D{\'{o}}ra, Cserti, and
  Cortijo}}]{Oroszlny2018}
\bibinfo{author}{\bibfnamefont{L.}~\bibnamefont{Oroszl{\'{a}}ny}},
  \bibinfo{author}{\bibfnamefont{B.}~\bibnamefont{D{\'{o}}ra}},
  \bibinfo{author}{\bibfnamefont{J.}~\bibnamefont{Cserti}}, \bibnamefont{and}
  \bibinfo{author}{\bibfnamefont{A.}~\bibnamefont{Cortijo}},
  \bibinfo{journal}{Physical Review B} \textbf{\bibinfo{volume}{97}},
  \bibinfo{pages}{205107} (\bibinfo{year}{2018}),
  \urlprefix\url{https://doi.org/10.1103/physrevb.97.205107}.

\bibitem[{\citenamefont{Mart{\'{\i}}n-Ruiz and Cortijo}(2018)}]{MartnRuiz2018}
\bibinfo{author}{\bibfnamefont{A.}~\bibnamefont{Mart{\'{\i}}n-Ruiz}}
  \bibnamefont{and} \bibinfo{author}{\bibfnamefont{A.}~\bibnamefont{Cortijo}},
  \bibinfo{journal}{Physical Review B} \textbf{\bibinfo{volume}{98}},
  \bibinfo{pages}{155125} (\bibinfo{year}{2018}),
  \urlprefix\url{https://doi.org/10.1103/physrevb.98.155125}.

\bibitem[{\citenamefont{D{\'{o}}ra and Moessner}(2010)}]{Dra2010}
\bibinfo{author}{\bibfnamefont{B.}~\bibnamefont{D{\'{o}}ra}} \bibnamefont{and}
  \bibinfo{author}{\bibfnamefont{R.}~\bibnamefont{Moessner}},
  \bibinfo{journal}{Physical Review B} \textbf{\bibinfo{volume}{81}},
  \bibinfo{pages}{165431} (\bibinfo{year}{2010}),
  \urlprefix\url{https://doi.org/10.1103/physrevb.81.165431}.

\bibitem[{\citenamefont{Vajna et~al.}(2015)\citenamefont{Vajna, D{\'{o}}ra, and
  Moessner}}]{Vajna2015}
\bibinfo{author}{\bibfnamefont{S.}~\bibnamefont{Vajna}},
  \bibinfo{author}{\bibfnamefont{B.}~\bibnamefont{D{\'{o}}ra}},
  \bibnamefont{and} \bibinfo{author}{\bibfnamefont{R.}~\bibnamefont{Moessner}},
  \bibinfo{journal}{Physical Review B} \textbf{\bibinfo{volume}{92}},
  \bibinfo{pages}{085122} (\bibinfo{year}{2015}),
  \urlprefix\url{https://doi.org/10.1103/physrevb.92.085122}.

\bibitem[{\citenamefont{Barati and Abedinpour}(2017)}]{Barati2017}
\bibinfo{author}{\bibfnamefont{S.}~\bibnamefont{Barati}} \bibnamefont{and}
  \bibinfo{author}{\bibfnamefont{S.~H.} \bibnamefont{Abedinpour}},
  \bibinfo{journal}{Physical Review B} \textbf{\bibinfo{volume}{96}},
  \bibinfo{pages}{155150} (\bibinfo{year}{2017}),
  \urlprefix\url{https://doi.org/10.1103/physrevb.96.155150}.

\bibitem[{\citenamefont{Singha et~al.}(2017)\citenamefont{Singha, Pariari,
  Satpati, and Mandal}}]{Singha2017}
\bibinfo{author}{\bibfnamefont{R.}~\bibnamefont{Singha}},
  \bibinfo{author}{\bibfnamefont{A.~K.} \bibnamefont{Pariari}},
  \bibinfo{author}{\bibfnamefont{B.}~\bibnamefont{Satpati}}, \bibnamefont{and}
  \bibinfo{author}{\bibfnamefont{P.}~\bibnamefont{Mandal}},
  \bibinfo{journal}{Proceedings of the National Academy of Sciences}
  \textbf{\bibinfo{volume}{114}}, \bibinfo{pages}{2468} (\bibinfo{year}{2017}),
  \urlprefix\url{https://doi.org/10.1073/pnas.1618004114}.

\bibitem[{\citenamefont{Rudenko and Yuan}(2020)}]{Rudenko2020}
\bibinfo{author}{\bibfnamefont{A.~N.} \bibnamefont{Rudenko}} \bibnamefont{and}
  \bibinfo{author}{\bibfnamefont{S.}~\bibnamefont{Yuan}},
  \bibinfo{journal}{Physical Review B} \textbf{\bibinfo{volume}{101}},
  \bibinfo{pages}{115127} (\bibinfo{year}{2020}),
  \urlprefix\url{https://doi.org/10.1103/physrevb.101.115127}.

\bibitem[{\citenamefont{Fu et~al.}(2019)\citenamefont{Fu, Yi, Zhang, Caputo,
  Ma, Gao, Lv, Kong, Huang, Richard et~al.}}]{Fu2019}
\bibinfo{author}{\bibfnamefont{B.-B.} \bibnamefont{Fu}},
  \bibinfo{author}{\bibfnamefont{C.-J.} \bibnamefont{Yi}},
  \bibinfo{author}{\bibfnamefont{T.-T.} \bibnamefont{Zhang}},
  \bibinfo{author}{\bibfnamefont{M.}~\bibnamefont{Caputo}},
  \bibinfo{author}{\bibfnamefont{J.-Z.} \bibnamefont{Ma}},
  \bibinfo{author}{\bibfnamefont{X.}~\bibnamefont{Gao}},
  \bibinfo{author}{\bibfnamefont{B.~Q.} \bibnamefont{Lv}},
  \bibinfo{author}{\bibfnamefont{L.-Y.} \bibnamefont{Kong}},
  \bibinfo{author}{\bibfnamefont{Y.-B.} \bibnamefont{Huang}},
  \bibinfo{author}{\bibfnamefont{P.}~\bibnamefont{Richard}},
  \bibnamefont{et~al.}, \bibinfo{journal}{Science Advances}
  \textbf{\bibinfo{volume}{5}}, \bibinfo{pages}{1126} (\bibinfo{year}{2019}),
  \urlprefix\url{https://doi.org/10.1126/sciadv.aau6459}.

\bibitem[{\citenamefont{Hosen et~al.}(2017)\citenamefont{Hosen, Dimitri,
  Belopolski, Maldonado, Sankar, Dhakal, Dhakal, Cole, Oppeneer, Kaczorowski
  et~al.}}]{Hosen2017}
\bibinfo{author}{\bibfnamefont{M.~M.} \bibnamefont{Hosen}},
  \bibinfo{author}{\bibfnamefont{K.}~\bibnamefont{Dimitri}},
  \bibinfo{author}{\bibfnamefont{I.}~\bibnamefont{Belopolski}},
  \bibinfo{author}{\bibfnamefont{P.}~\bibnamefont{Maldonado}},
  \bibinfo{author}{\bibfnamefont{R.}~\bibnamefont{Sankar}},
  \bibinfo{author}{\bibfnamefont{N.}~\bibnamefont{Dhakal}},
  \bibinfo{author}{\bibfnamefont{G.}~\bibnamefont{Dhakal}},
  \bibinfo{author}{\bibfnamefont{T.}~\bibnamefont{Cole}},
  \bibinfo{author}{\bibfnamefont{P.~M.} \bibnamefont{Oppeneer}},
  \bibinfo{author}{\bibfnamefont{D.}~\bibnamefont{Kaczorowski}},
  \bibnamefont{et~al.}, \bibinfo{journal}{Physical Review B}
  \textbf{\bibinfo{volume}{95}}, \bibinfo{pages}{161101(R)}
  (\bibinfo{year}{2017}),
  \urlprefix\url{https://doi.org/10.1103/physrevb.95.161101}.

\bibitem[{\citenamefont{Vitanov and Garraway}(1996)}]{Vitanov1996}
\bibinfo{author}{\bibfnamefont{N.~V.} \bibnamefont{Vitanov}} \bibnamefont{and}
  \bibinfo{author}{\bibfnamefont{B.~M.} \bibnamefont{Garraway}},
  \bibinfo{journal}{Physical Review A} \textbf{\bibinfo{volume}{53}},
  \bibinfo{pages}{4288} (\bibinfo{year}{1996}),
  \urlprefix\url{https://doi.org/10.1103/physreva.53.4288}.

\bibitem[{\citenamefont{Cohen and McGady}(2008)}]{Cohen2008}
\bibinfo{author}{\bibfnamefont{T.~D.} \bibnamefont{Cohen}} \bibnamefont{and}
  \bibinfo{author}{\bibfnamefont{D.~A.} \bibnamefont{McGady}},
  \bibinfo{journal}{Physical Review D} \textbf{\bibinfo{volume}{78}},
  \bibinfo{pages}{036008} (\bibinfo{year}{2008}),
  \urlprefix\url{https://doi.org/10.1103/physrevd.78.036008}.

\bibitem[{\citenamefont{Lehto and Suominen}(2012)}]{Lehto2012}
\bibinfo{author}{\bibfnamefont{J.}~\bibnamefont{Lehto}} \bibnamefont{and}
  \bibinfo{author}{\bibfnamefont{K.-A.} \bibnamefont{Suominen}},
  \bibinfo{journal}{Physical Review A} \textbf{\bibinfo{volume}{86}},
  \bibinfo{pages}{033415} (\bibinfo{year}{2012}),
  \urlprefix\url{https://doi.org/10.1103/physreva.86.033415}.

\bibitem[{\citenamefont{Tanji}(2009)}]{Tanji2009}
\bibinfo{author}{\bibfnamefont{N.}~\bibnamefont{Tanji}},
  \bibinfo{journal}{Annals of Physics} \textbf{\bibinfo{volume}{324}},
  \bibinfo{pages}{1691} (\bibinfo{year}{2009}),
  \urlprefix\url{https://doi.org/10.1016/j.aop.2009.03.012}.

\bibitem[{\citenamefont{Ashcroft}(1976)}]{Ashcroft1976}
\bibinfo{author}{\bibfnamefont{N.~W.} \bibnamefont{Ashcroft}},
  \emph{\bibinfo{title}{Solid State Physics}} (\bibinfo{publisher}{Cengage
  Learning, Inc}, \bibinfo{year}{1976}), ISBN \bibinfo{isbn}{0030839939}.

\bibitem[{\citenamefont{Boross et~al.}(2011)\citenamefont{Boross, D{\'{o}}ra,
  and Moessner}}]{Boross2011}
\bibinfo{author}{\bibfnamefont{P.}~\bibnamefont{Boross}},
  \bibinfo{author}{\bibfnamefont{B.}~\bibnamefont{D{\'{o}}ra}},
  \bibnamefont{and} \bibinfo{author}{\bibfnamefont{R.}~\bibnamefont{Moessner}},
  \bibinfo{journal}{physica status solidi (b)} \textbf{\bibinfo{volume}{248}},
  \bibinfo{pages}{2627} (\bibinfo{year}{2011}),
  \urlprefix\url{https://doi.org/10.1002/pssb.201100184}.

\bibitem[{\citenamefont{Gavrilov and Gitman}(1996)}]{Gavrilov1996}
\bibinfo{author}{\bibfnamefont{S.~P.} \bibnamefont{Gavrilov}} \bibnamefont{and}
  \bibinfo{author}{\bibfnamefont{D.~M.} \bibnamefont{Gitman}},
  \bibinfo{journal}{Physical Review D} \textbf{\bibinfo{volume}{53}},
  \bibinfo{pages}{7162} (\bibinfo{year}{1996}),
  \urlprefix\url{https://doi.org/10.1103/physrevd.53.7162}.

\bibitem[{\citenamefont{Casher et~al.}(1979)\citenamefont{Casher, Neuberger,
  and Nussinov}}]{Casher1979}
\bibinfo{author}{\bibfnamefont{A.}~\bibnamefont{Casher}},
  \bibinfo{author}{\bibfnamefont{H.}~\bibnamefont{Neuberger}},
  \bibnamefont{and} \bibinfo{author}{\bibfnamefont{S.}~\bibnamefont{Nussinov}},
  \bibinfo{journal}{Physical Review D} \textbf{\bibinfo{volume}{20}},
  \bibinfo{pages}{179} (\bibinfo{year}{1979}),
  \urlprefix\url{https://doi.org/10.1103/physrevd.20.179}.

\bibitem[{\citenamefont{Suominen}(1992)}]{Suominen1992}
\bibinfo{author}{\bibfnamefont{K.-A.} \bibnamefont{Suominen}},
  \bibinfo{journal}{Optics Communications} \textbf{\bibinfo{volume}{93}},
  \bibinfo{pages}{126} (\bibinfo{year}{1992}),
  \urlprefix\url{https://doi.org/10.1016/0030-4018(92)90140-m}.

\bibitem[{\citenamefont{Schwinger}(1951)}]{Schwinger1951}
\bibinfo{author}{\bibfnamefont{J.}~\bibnamefont{Schwinger}},
  \bibinfo{journal}{Physical Review} \textbf{\bibinfo{volume}{82}},
  \bibinfo{pages}{664} (\bibinfo{year}{1951}),
  \urlprefix\url{https://doi.org/10.1103/physrev.82.664}.

\bibitem[{\citenamefont{Zwillinger}(2014)}]{Gradstein2014}
\bibinfo{author}{\bibfnamefont{D.}~\bibnamefont{Zwillinger}},
  \emph{\bibinfo{title}{Table of Integrals, Series, and Products}}
  (\bibinfo{publisher}{Academic Press}, \bibinfo{year}{2014}),
  \bibinfo{edition}{eighth} ed.

\bibitem[{\citenamefont{Davis}(1976)}]{Davis1976}
\bibinfo{author}{\bibfnamefont{J.~P.} \bibnamefont{Davis}},
  \bibinfo{journal}{The Journal of Chemical Physics}
  \textbf{\bibinfo{volume}{64}}, \bibinfo{pages}{3129} (\bibinfo{year}{1976}),
  \urlprefix\url{https://doi.org/10.1063/1.432648}.

\bibitem[{\citenamefont{Dykhne}(1962)}]{Dykhne1962}
\bibinfo{author}{\bibfnamefont{A.}~\bibnamefont{Dykhne}},
  \bibinfo{journal}{JTEP} \textbf{\bibinfo{volume}{41}}, \bibinfo{pages}{1324}
  (\bibinfo{year}{1962}).

\bibitem[{\citenamefont{Kazantsev et~al.}(1990)\citenamefont{Kazantsev,
  Surdutovich, and Yakovlev}}]{Kazantsev1990}
\bibinfo{author}{\bibfnamefont{A.~P.} \bibnamefont{Kazantsev}},
  \bibinfo{author}{\bibfnamefont{G.~I.} \bibnamefont{Surdutovich}},
  \bibnamefont{and} \bibinfo{author}{\bibfnamefont{V.~P.}
  \bibnamefont{Yakovlev}}, \emph{\bibinfo{title}{Mechanical Action of Light on
  Atoms}} (\bibinfo{publisher}{{WORLD} {SCIENTIFIC}}, \bibinfo{year}{1990}),
  \urlprefix\url{https://doi.org/10.1142/0585}.

\bibitem[{\citenamefont{Gu et~al.}(2019)\citenamefont{Gu, Liu, Mei, Jia, Zhang,
  and Xue}}]{Gu2019}
\bibinfo{author}{\bibfnamefont{F.-L.} \bibnamefont{Gu}},
  \bibinfo{author}{\bibfnamefont{J.}~\bibnamefont{Liu}},
  \bibinfo{author}{\bibfnamefont{F.}~\bibnamefont{Mei}},
  \bibinfo{author}{\bibfnamefont{S.}~\bibnamefont{Jia}},
  \bibinfo{author}{\bibfnamefont{D.-W.} \bibnamefont{Zhang}}, \bibnamefont{and}
  \bibinfo{author}{\bibfnamefont{Z.-Y.} \bibnamefont{Xue}},
  \bibinfo{journal}{npj Quantum Information} \textbf{\bibinfo{volume}{5}},
  \bibinfo{pages}{36} (\bibinfo{year}{2019}),
  \urlprefix\url{https://doi.org/10.1038/s41534-019-0148-9}.

\bibitem[{\citenamefont{Tarruell et~al.}(2012)\citenamefont{Tarruell, Greif,
  Uehlinger, Jotzu, and Esslinger}}]{Tarruell2012}
\bibinfo{author}{\bibfnamefont{L.}~\bibnamefont{Tarruell}},
  \bibinfo{author}{\bibfnamefont{D.}~\bibnamefont{Greif}},
  \bibinfo{author}{\bibfnamefont{T.}~\bibnamefont{Uehlinger}},
  \bibinfo{author}{\bibfnamefont{G.}~\bibnamefont{Jotzu}}, \bibnamefont{and}
  \bibinfo{author}{\bibfnamefont{T.}~\bibnamefont{Esslinger}},
  \bibinfo{journal}{Nature} \textbf{\bibinfo{volume}{483}},
  \bibinfo{pages}{7389} (\bibinfo{year}{2012}),
  \urlprefix\url{https://doi.org/10.1038/nature10871}.

\bibitem[{\citenamefont{Song et~al.}(2019)\citenamefont{Song, He, Niu, Zhang,
  Ren, Liu, and Jo}}]{Song2019}
\bibinfo{author}{\bibfnamefont{B.}~\bibnamefont{Song}},
  \bibinfo{author}{\bibfnamefont{C.}~\bibnamefont{He}},
  \bibinfo{author}{\bibfnamefont{S.}~\bibnamefont{Niu}},
  \bibinfo{author}{\bibfnamefont{L.}~\bibnamefont{Zhang}},
  \bibinfo{author}{\bibfnamefont{Z.}~\bibnamefont{Ren}},
  \bibinfo{author}{\bibfnamefont{X.-J.} \bibnamefont{Liu}}, \bibnamefont{and}
  \bibinfo{author}{\bibfnamefont{G.-B.} \bibnamefont{Jo}},
  \bibinfo{journal}{Nature Physics} \textbf{\bibinfo{volume}{15}},
  \bibinfo{pages}{911} (\bibinfo{year}{2019}),
  \urlprefix\url{https://doi.org/10.1038/s41567-019-0564-y}.

\bibitem[{\citenamefont{BenDahan et~al.}(1996)\citenamefont{BenDahan, Peik,
  Reichel, Castin, and Salomon}}]{BenDahan1996}
\bibinfo{author}{\bibfnamefont{M.}~\bibnamefont{BenDahan}},
  \bibinfo{author}{\bibfnamefont{E.}~\bibnamefont{Peik}},
  \bibinfo{author}{\bibfnamefont{J.}~\bibnamefont{Reichel}},
  \bibinfo{author}{\bibfnamefont{Y.}~\bibnamefont{Castin}}, \bibnamefont{and}
  \bibinfo{author}{\bibfnamefont{C.}~\bibnamefont{Salomon}},
  \bibinfo{journal}{Physical Review Letters} \textbf{\bibinfo{volume}{76}},
  \bibinfo{pages}{4508} (\bibinfo{year}{1996}),
  \urlprefix\url{https://doi.org/10.1103/physrevlett.76.4508}.

\bibitem[{\citenamefont{Novak et~al.}(2019)\citenamefont{Novak, Zhang,
  Orbani{\'{c}}, Bili{\v{s}}kov, Eguchi, Paschen, Kimura, Wang, Osada, Uchida
  et~al.}}]{Novak2019}
\bibinfo{author}{\bibfnamefont{M.}~\bibnamefont{Novak}},
  \bibinfo{author}{\bibfnamefont{S.~N.} \bibnamefont{Zhang}},
  \bibinfo{author}{\bibfnamefont{F.}~\bibnamefont{Orbani{\'{c}}}},
  \bibinfo{author}{\bibfnamefont{N.}~\bibnamefont{Bili{\v{s}}kov}},
  \bibinfo{author}{\bibfnamefont{G.}~\bibnamefont{Eguchi}},
  \bibinfo{author}{\bibfnamefont{S.}~\bibnamefont{Paschen}},
  \bibinfo{author}{\bibfnamefont{A.}~\bibnamefont{Kimura}},
  \bibinfo{author}{\bibfnamefont{X.~X.} \bibnamefont{Wang}},
  \bibinfo{author}{\bibfnamefont{T.}~\bibnamefont{Osada}},
  \bibinfo{author}{\bibfnamefont{K.}~\bibnamefont{Uchida}},
  \bibnamefont{et~al.}, \bibinfo{journal}{Physical Review B}
  \textbf{\bibinfo{volume}{100}}, \bibinfo{pages}{085137}
  (\bibinfo{year}{2019}),
  \urlprefix\url{https://doi.org/10.1103/physrevb.100.085137}.

\end{thebibliography}
	
\end{document}